\documentclass[useAMS,usenatbib]{mnras}

\usepackage[T1]{fontenc}
\usepackage[utf8]{inputenc}
\usepackage{times}
\usepackage{amsmath}
\usepackage{amssymb}
\usepackage{bm}
\usepackage[english]{babel}
\usepackage{graphicx}
\usepackage{color,xcolor}
\usepackage{placeins}
\usepackage{array}

\colorlet{red}{black}

\title[]{Magnetization of the intergalactic medium in the IllustrisTNG simulations: the importance of extended, outflow-driven bubbles}
\author[]{Andrés Arámburo-García$^{1}$\thanks{aramburo@strw.leidenuniv.nl}, Kyrylo Bondarenko$^{2,3}$\thanks{kyrylo.bondarenko@cern.ch}, Alexey Boyarsky$^{1}$\thanks{boyarsky@lorentz.leidenuniv.nl}, \newauthor
Dylan Nelson$^{4}$\thanks{dnelson@uni-heidelberg.de},
Annalisa Pillepich$^{5}$\thanks{pillepich@mpia-hd.mpg.de}, and Anastasia Sokolenko$^{6}$\thanks{anastasia.sokolenko@oeaw.ac.at}
  \\ $^1$ Institute  Lorentz, Leiden University, Niels Bohrweg 2, Leiden, NL-2333 CA, the Netherlands\\
  $^{2}$Theoretical Physics Department, CERN, 
 Geneva 23, CH-1211, Switzerland \\
 $^3$ L'Ecole polytechnique f\'ed\'erale de Lausanne, 1015 Lausanne, Switzerland 
 \\ $^4$ Universität Heidelberg, Zentrum für Astronomie, Institut für theoretische Astrophysik, Albert-Ueberle-Str. 2, 69120 Heidelberg, Germany \\
  $^{5}$ Max-Planck-Institut f\"{u}r Astronomie, K\"{o}nigstuhl 17, 69117 Heidelberg, Germany\\
 $^6$ Institute of High Energy Physics, Austrian Academy of Sciences, Nikolsdorfergasse 18, 1050 Vienna, Austria}

\begin{document}
\maketitle
\begin{abstract}
We study the effects of galaxy formation physics on the magnetization of the intergalactic medium (IGM) using the IllustrisTNG simulations. We demonstrate that large-scale regions affected by the outflows from galaxies and clusters contain magnetic fields that are several orders of magnitude stronger than in unaffected regions with the same electron density. 
Moreover, like magnetic fields amplified inside galaxies, these magnetic fields do not depend on the primordial seed, i.e. the adopted initial conditions for magnetic field strength.
We study the volume filling fraction of these strong field regions and their occurrence in random lines of sight. As a first application, we use these results to put bounds on the photon-axion conversion from spectral distortion of the CMB. 
As photon-axion coupling grows with energy, stronger constraints could potentially be obtained using data on the propagation of gamma-ray photons through the IGM.  Finally, we also briefly discuss potential applications of our results to the Faraday Rotation measurements.
\end{abstract}


\section{Introduction}
The Universe is magnetized on all scales -- from planets and stars to galaxy clusters and beyond. Magnetic fields affect the propagation of charged particles and therefore play an important role in the physics of the Earth's atmosphere, the Sun and solar system, galaxies, and so on. However, the origin of magnetic fields in galaxies and, especially, galaxy clusters in the low-redshift Universe remains an important open problem (see e.g.~\citealt{Durrer:2013pga} for a review). 

Indeed, there are two stages of magnetogenesis -- the first stage is the generation of \textcolor{red}{weak} seed magnetic fields and the second stage is their subsequent evolution during structure formation. Seed fields could have either primordial (produced before recombination) or astrophysical origin (produced after the formation of the first stars), see e.g.~\cite{Subramanian:2015lua} for a review. Primordial seed fields fill the whole universe, although they are not necessarily constant, their correlation length depends on the production mechanism and the epoch when they were produced. 
\textcolor{red}{An} astrophysical seed magnetic field is generated by Biermann battery-type mechanisms when the curl of the electric field is created by non-parallel gradients of density and temperature, giving rise to a magnetic field via Faraday induction~\citep{Subramanian:2015lua}. Such a mechanism could generate magnetic fields during star formation, reionization~\citep{Subramanian1994,Gnedin:2000ax}, or even during the later collapse of galaxies and halos in cosmological shocks~\citep{Kulsrud:1996km}. For example,~\cite{Vazza:2017qge} \textcolor{red}{simulate} 25 different scenarios of initial magnetogenesis, both primordial and astrophysical. 

In the second stage, magnetic fields evolve with the expansion of the Universe and structure formation. Outside structures, these fields dilute approximately as $\sim a^{-2}$~\citep{Durrer:2013pga}, \textcolor{red}{where $a$ is the scale factor}. In the regions where dense structures form, magnetic fields are adiabatically compressed and, moreover, strongly amplified (up to several $\mu$G~\citealt{Pakmor:2013rqa,Rieder2017}) by different dynamo mechanisms driven by the baryonic physics of galaxy formation~\citep{1955ApJ...122..293P,1988Natur.336..341R,1999ARA&A..37...37K,Brandenburg:2004jv,Kulsrud:2007an,2016MNRAS.457.1722R,2017MNRAS.471..144S,2017ApJ...843..113B,2019MNRAS.483.1008S,2018MNRAS.479.3343M,2018MNRAS.480.3907V}. Amplification in filaments also occurs via shear flows~\citep{1999ApJ...518..177B,1999A&A...348..351D,Dolag:2004kp}. As a result, magnetic fields in galaxies and collapsed structures, amplified by many orders of magnitude by gravo-magnetohydrodynamics (MHD) dynamos, ``forget'' the properties of the initial magnetic fields~\citep[\textcolor{red}{see e.g.}][\textcolor{red}{for cosmological simulations of galaxies}]{Pakmor:2013rqa,Marinacci:2015dja,Pillepich2018MNRAS.473.4077P}. To the contrary, magnetic fields that are far from structures are much closer to the diluted initial fields and \textcolor{red}{could therefore be used to infer information about the properties and the origin of the initial fields.}

On the observational side, cosmic magnetic fields are relatively well studied in virialized objects like galaxies and galaxy clusters. A powerful method to detect these fields is Faraday Rotation Measure (RM) see e.g.~\cite{Brentjens:2005zc} and references therein. With the current generation of instruments, this method is efficient for magnetic fields with the strength of $B \ge\mathcal{O}(1)$~nG~\citep{Durrer:2013pga}. Such magnetic fields exist mainly in the dense centers of collapsed structures -- galaxies, galaxy groups, and clusters (e.g.~\citealt{Carilli:2001hj,Laing:2008yf,2015A&A...578A..93B,vanWeeren:2019vxy}). However, these objects fill only a small fraction of the volume of the Universe.

Empirical constraints on the cosmic magnetic fields outside galaxies and clusters remain difficult. Attempts to measure magnetic fields in filaments by Faraday Rotation Measure with LOFAR~\citep{OSullivan:2020pll} place only an upper bound of a few nG, on Mpc scales, consistent with other \textcolor{red}{works}~\citep{Ravi:2016kfj,Vernstrom:2019gjr,Blasi:1999hu,Pshirkov:2015tua,10.1093/mnras/stw1903,Bray:2018ipq}. In addition, a lower bound can be obtained with high-energy gamma-ray data~\citep{Neronov:1900zz,Dermer:2010mm,Tavecchio:2010ja,Dolag:2010ni,Taylor:2011bn}. This lower bound is as weak as $B\gtrsim 10^{-17}$~G on Mpc scales. 
In the future, the upper bound can be improved with next-generation radio telescopes such as the Square Kilometer Array (SKA)~\citep{Carilli:2004nx}, while the lower bound may be improved by high-energy observatories including CTA~\citep{CTAConsortium:2018tzg}, which is expected to start obtaining data soon. For now, the observational uncertainty in the properties of the intergalactic magnetic fields (IGMF) outside galactic halos is large. 

From the theoretical as well as observational perspective our knowledge on the magnetic fields outside galaxies and clusters is rather limited. At the same time, magnetic fields in the large volumes of the intergalactic medium (IGM) can play a profound role in many important problems in physics. For example, magnetic fields can strongly affect the propagation of light and the spectra of various astrophysical sources (see e.g. ~\citealt{Brentjens:2005zc,Neronov:1900zz}) as well as propagation of cosmic rays~\citep{AlvesBatista:2017vob}. 
If the magnetic field along the lines of sight to certain classes of sources were better constrained, these effects (see examples below) can give us important insight into fundamental physics. 

When light propagates towards us from remote sources, most of the intervening pathlength is typically not in virialized objects, as these occupy only a tiny fraction of the Universe by volume. Rather, photons propagate through the IGM -- the space between dark matter halos occupied by less dense regions including cosmic voids, sheets, and filamentary structures. Both the free electron number density and magnetic field strength in the IGM evolve with time, and theoretical modeling of this evolution, through the epoch of reionization and down to the present day, remains a challenge. Therefore, to probe the effects of cosmic magnetic fields on light propagation it is not enough to model magnetic fields only within halos.

One regime where magnetic fields in the IGM play a crucial role \textcolor{red}{is} gamma-ray astronomy. The photons from high-energy gamma-ray sources create electron-positron pairs interacting with the extra-galactic background light. These charged particles can then emit secondary gamma-ray photons interacting with the CMB via the inverse Compton effect. The presence of a magnetic field results in a deviation of the charged particles and, therefore, in a change in the morphology of the signal (see e.g.~\citealt{Neronov:2009gh} for a more detailed discussion).\footnote{This effect places constraints on the IGMF from gamma-ray observations, as mentioned above.}

Magnetic fields can also affect light propagation in the presence of new, as of yet unobserved particles that are not included in the Standard Model of particle physics. A famous example is an axion or axion-like particle (ALP), initially introduced to explain why CP violation in QCD is so tiny~\citep{Weinberg:1977ma,Wilczek:1977pj}. Axions have been theorized to play the role of dark matter~\citep{Preskill:1982cy}. Photons can be converted into ALPs, but only when they pass through magnetized regions of the Universe. In this case, the conversion probability depends sensitively on the strength of the magnetic field~\citep{1983PhRvL..51.1415S,1988PhRvD..37.1237R}.

In this paper, we use the IllustrisTNG suite of cosmological simulations \textcolor{red}{(see Section~\ref{sec:sim} for details)}, as well as additional variation runs performed with different values of the initial magnetic fields and with and without feedback, to study the regions of the IGM that could be affected by galactic outflows. The starting point of our investigation is the idea that the strong magnetic fields generated deep within dark matter halos can affect and extend to much larger volumes, as baryonic outflows, caused by strong feedback processes, eject magnetized gas to regions extending far beyond halo scales \textcolor{red}{(and not just beyond galactic scales, as studied e.g. in~\citealt{marinacci18,2020MNRAS.494.4393S,2020MNRAS.498.3125P,2010A&A...523A..72D,2020MNRAS.495.4475M}).} 

In this work, we concentrate our attention on the regions of the IGM affected by galactic outflows, their origin, and their impact on magnetic fields in the IGM. We discuss the properties and strength of the IGMF predicted in IllustrisTNG (Section~\ref{sec:sim}), presenting our results in a form that may be used for different applications. We show that the magnetic fields affected by galactic outflows depend more on the MHD processes occurring within galaxies rather than on the primordial magnetic field seeds. In particular, we show that the predicted magnetic field strength in the regions affected by outflows is similar for runs spanning orders of magnitude different strength of the initial magnetic fields \textcolor{red}{(Section~\ref{sec:igm})}. At the same time, magnetic fields in these regions are orders of magnitude larger than in other regions of the IGM with similar matter density. As a first example application, we apply our findings to constrain ALPs (Section~\ref{sec:axion}) and summarize our results (Section~\ref{sec:conclusion}).

\begin{figure*}
    \centering
    \includegraphics[width=1\textwidth]{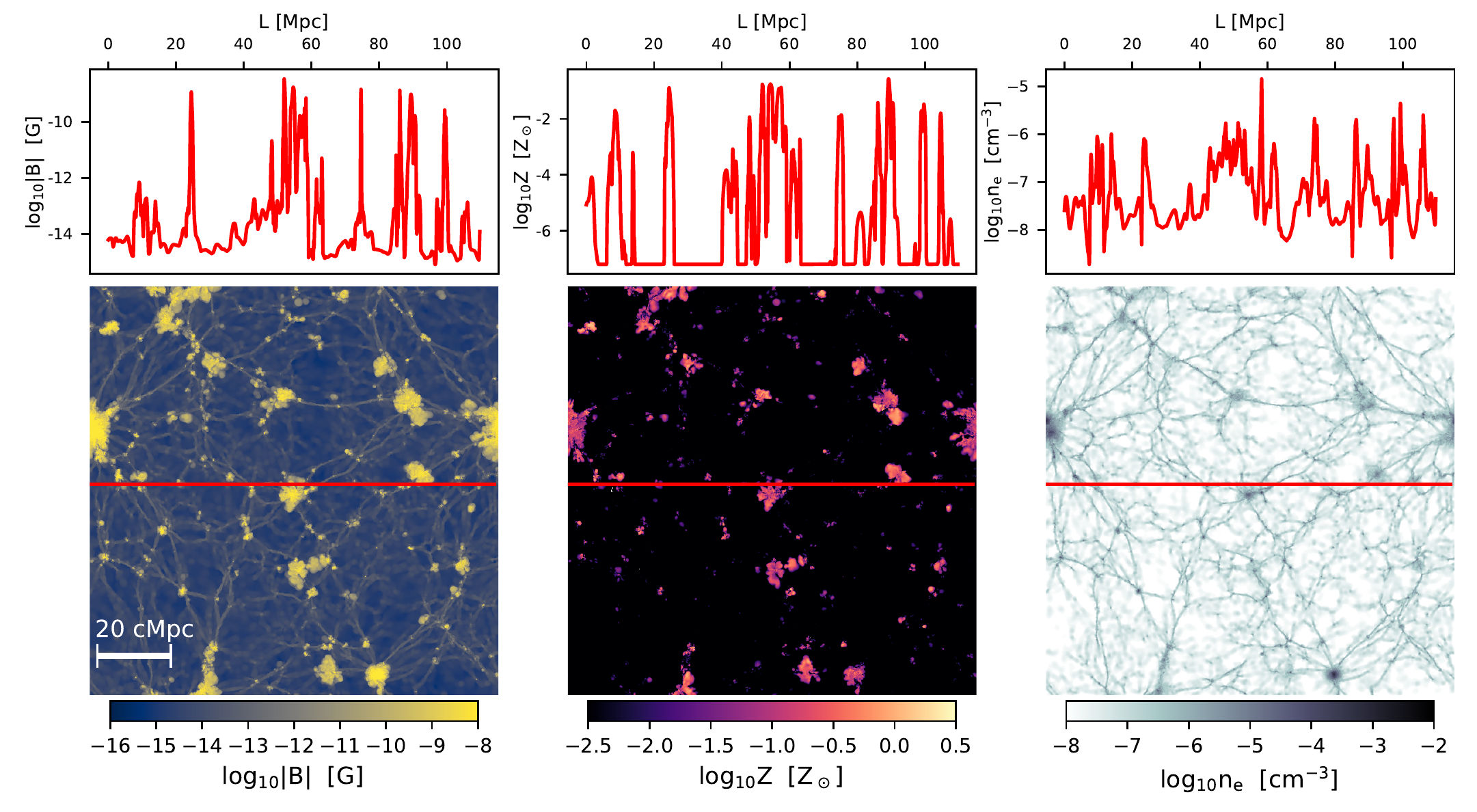}
    \caption{Lower panels show distributions of magnetic field, metallicity and electron number density for a thin slice of the the whole $z=0$ TNG100 box, $(110\text{ Mpc})^2 \times 20$~kpc. The upper panels show values of corresponding quantities along one particular line of sight (indicated by the red line on the lower panels). Significant variation occurs as sightlines pass through underdense versus overdense regions.
    }
    \label{fig:Bslice}
\end{figure*}

\section{Methods}
\label{sec:sim}

\subsection{Simulations}

IllustrisTNG (TNG) is a suite of large-volume cosmological gravo-magnetohydrodynamic simulations~\citep{nelson18,springel18,pillepich18,naiman18,marinacci18}. Each evolves initial conditions from $z=127$ to the present time, following the evolution of gas, stars, and black holes (baryons), together with dark matter. The TNG simulations use the moving-mesh \textsc{Arepo} code~\citep{springel2010MNRAS.401..791S} to solve the coupled equations of self-gravity and ideal MHD~\citep{Pakmor2011MNRAS.418.1392P,Pakmor2013MNRAS.432..176P}, and adopt cosmological parameters consistent with Planck 2015~\citep{Plank2016A&A...594A..13P}. The simulations include a comprehensive galaxy formation model incorporating astrophysical processes such as gas metal-line cooling and heating, star formation, stellar evolution, and heavy element enrichment, supermassive black hole growth, AGN feedback, and galactic winds launched by supernovae~\citep{Weinberger2017MNRAS.465.3291W,Pillepich2018MNRAS.473.4077P}. The TNG project currently spans three different volumes, TNG50, TNG100, and TNG300, each run with several different numerical resolutions. In this work, we mainly use the publicly available TNG100-1 simulation~\citep{Nelson2019ComAC...6....2N}, the highest resolution run of TNG100, with a box side-length of $L \sim 100~\text{cMpc}$ \textcolor{red}{(comoving Mpc)}. Containing $1820^3$ dark matter particles and an equal number of gas cells, it has a mass resolution of $m_{\text{bar}} = 1.4 \times 10^6~ \mathrm{M_\odot}$, and $m_{\text{DM}} = 7.5 \times 10^6~\mathrm{M_\odot}$, respectively. \textcolor{red}{From now on, we refer to such simulation as TNG100.}

\subsection{Galaxy formation model}

As we are particularly interested in the role of galactic-scale outflows in producing extended regions of high magnetic field strength, we describe the feedback physics briefly. A supermassive black hole (SMBH) is created in all dark matter halos which exceed a total mass of $\sim 7 \times 10^{10}$\,M$_{\odot}$, by placing a SMBH at the potential minimum with an initial mass of $\sim 10^{6}$\,M$_{\odot}$. These black holes subsequently grow via binary mergers with each other during galaxy collisions, and via smooth gas accretion from the surrounding environment. Black hole accretion is calculated using the Bondi-Hoyle-Lyttleton assumption~\citep{Weinberger2017MNRAS.465.3291W}, which depends on the black hole mass, local gas density, and relative velocity between the black hole and its surroundings. The accretion rate onto SMBHs is limited to the Eddington rate. To model energetic feedback from SMBHs, a small fraction of the rest mass of accreted matter is available to be deposited back into the locally surrounding gas. This energy is injected in a dual-mode model, depending on this accretion rate: one mode is for the high-accretion state (above $\sim 10$ percent of the Eddington rate), while the second, low-accretion state feedback mode, operates for accretion rates roughly below this value.

\begin{figure*}
    \centering
    \includegraphics[width=0.475\textwidth]{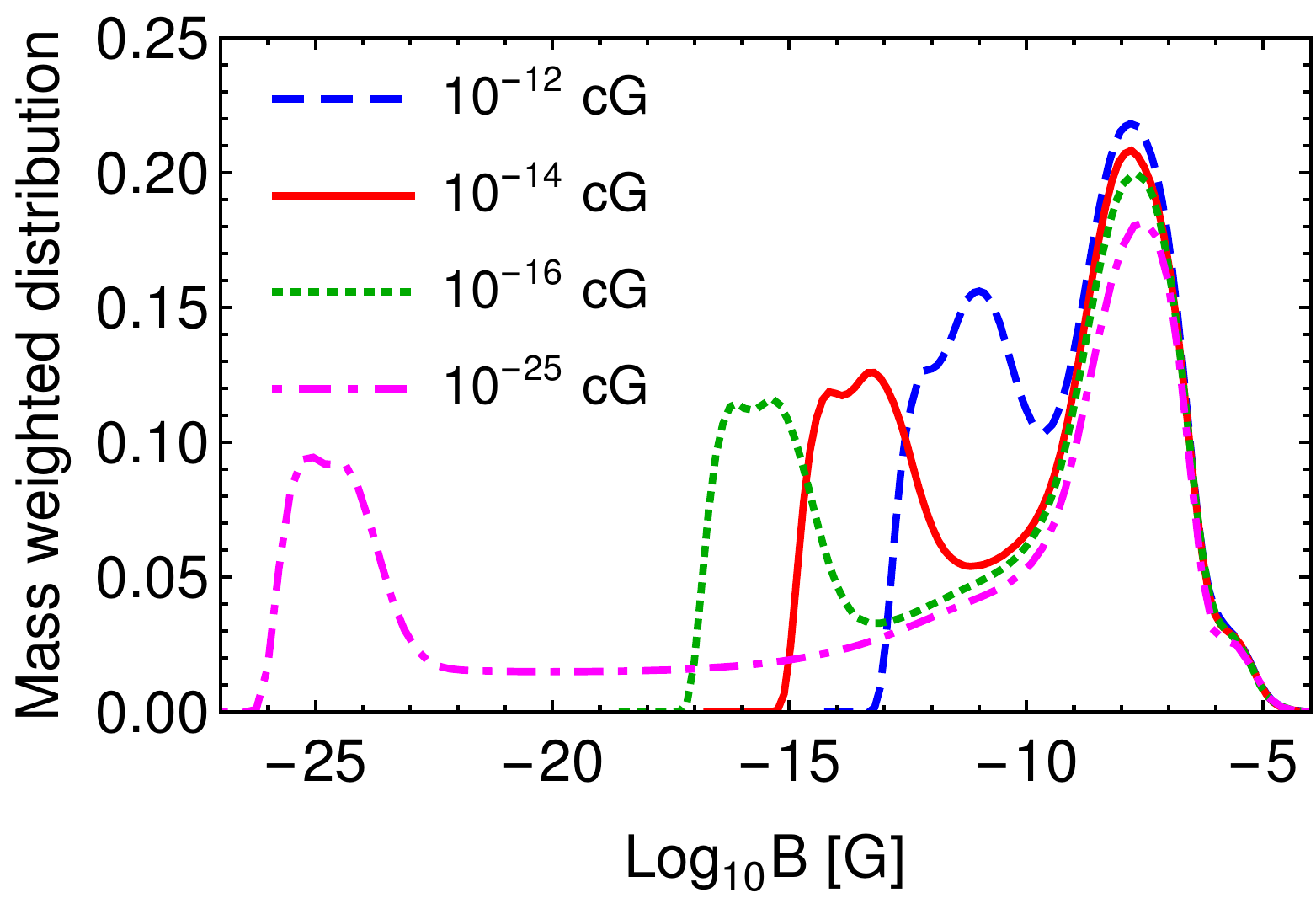}~
    \includegraphics[width=0.485\textwidth]{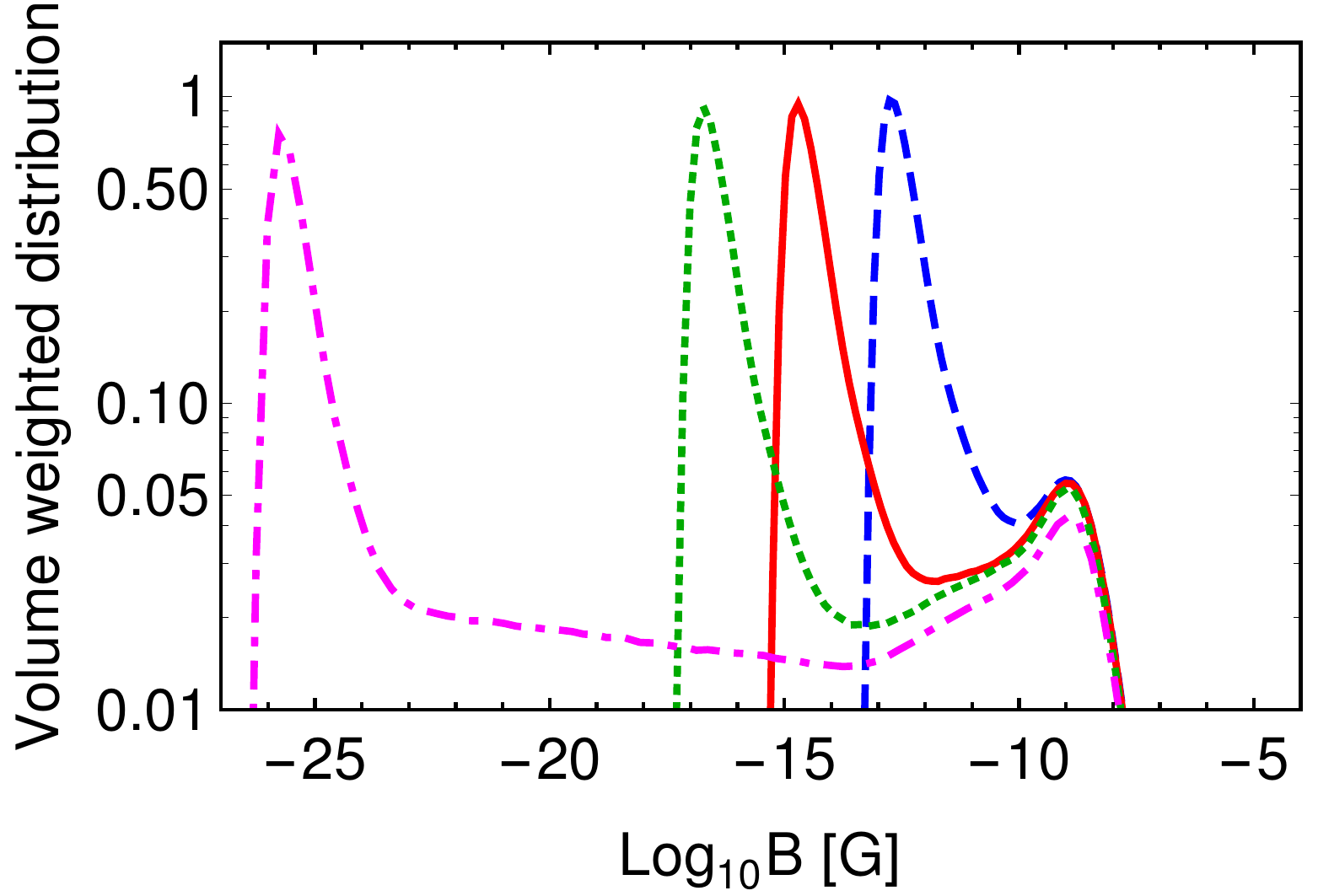}
    \caption{Mass weighted (left panel) and volume weighted (right panel) distributions of the magnetic field strength at $z=0$ for \textcolor{red}{four} simulations with different seed field values \textcolor{red}{but with the same underlying TNG galaxy formation model, over 25~cMpc/h a-side volumes: $10^{-12}$~cG (blue dashed), $10^{-14}$~cG (red solid, the fiducial choice of the TNG simulation adopted throughout the paper), $10^{-16}$~cG (green dotted), $10^{-25}$~cG (magenta dot-dashed)}, where cG are comoving Gauss. While the magnitude $|B|$ depends on the assumed initial field strength for very low field strengths, corresponding to underdense/void-like regions of the simulated universe \textcolor{red}{(see Fig.~\ref{fig:ne_B})}, the magnetic field strength within collapsed structures at $\log{(|B| / \rm{G})} \gtrsim -10$ is largely unchanged, due to the rapid amplification processes which effectively erase knowledge of the primordial seed fields \textcolor{red}{explored in this work}.}
    \label{fig:B3}
\end{figure*}

\begin{table*}
\setlength\extrarowheight{2pt}
\begin{tabular}{l|c|c|c|c||c|c|c|c|}
\cline{2-9}
 & \multicolumn{4}{c||}{Mass weighted fraction}        & \multicolumn{4}{c|}{Volume weighted fraction}      \\ \hline
\multicolumn{1}{|l|}{$B_0$ [cG]}      & $10^{-12}$ & $10^{-14}$ & $10^{-16}$ & $10^{-25}$ & $10^{-12}$ & $10^{-14}$ & $10^{-16}$ & $10^{-25}$ \\ \hline
\multicolumn{1}{|l|}{$B>10^{-12}$~cG} & $-$        & $68.6\%$   & $64.6\%$   & $57.5\%$   & $-$        & $14.8\%$   & $13.8\%$   & $11.4\%$   \\ \hline
\multicolumn{1}{|l|}{$B>10^{-11}$~cG} & $-$        & $62.8\%$   & $60.2\%$   & $53.6\%$   & $-$        & $12.1\%$   & $11.4\%$   & $9.5\%$    \\ \hline
\multicolumn{1}{|l|}{$B>10^{-10}$~cG} & $62.0\%$   & $57.0\%$   & $54.8\%$   & $48.9\%$   & $9.4\%$    & $9.0\%$    & $8.5\%$    & $7.1\%$    \\ \hline
\multicolumn{1}{|l|}{$B>10^{-9}$~cG}  & $50.9\%$   & $48.3\%$   & $46.7\%$   & $42.0\%$   & $4.6\%$    & $4.6\%$    & $4.3\%$    & $3.6\%$    \\ \hline
\end{tabular}
\caption{\textcolor{red}{Fraction of large-value magnetic fields calculated by mass and volume weighted distributions from \textcolor{red}{the 25 cMpc/h} simulations with different seed magnetic field values $B_0$ presented in Fig.~\ref{fig:B3}. We excluded values for two fractions for the largest seed magnetic field $10^{-12}$~cG because they are significantly contaminated by the seed field.}}
\label{tab:filling-fractions}
\end{table*}

At high accretion rates, energy is deposited in a continuous manner, by thermally heating gas. At low accretion rates, kinetic energy is injected in a discrete rather than continuous fashion, such that feedback events occur once enough energy accumulates (see \citealt{Weinberger2017MNRAS.465.3291W} for additional details). In this mode each injection event is modeled as a high-velocity kinetic wind, which is oriented randomly for each event, producing a time-average isotropic energy injection. This sub-resolution model is based on theoretical expectations for low accretion rate black holes, i.e. below one percent of Eddington, which develop radiatively inefficient flows and thereby convert gravitational binding energy into a non-relativistic wind~\citep{blandford1999,yuan2014}. In the TNG model, it is this population of low luminosity, slowly accreting SMBHs which drive the most powerful outflows~\citep{nelson2019b}. This occurs above a characteristic galaxy stellar mass (dark matter halo mass) of $\sim 10^{10.5}$\,M$_{\odot}$ ($\sim 10^{12}$\,M$_{\odot}$), corresponding to the onset of quenching in the galaxy population~\citep{weinberger2018,2020MNRAS.tmp.2921D}. Lower mass galaxies can also produce strong outflows, via a model for supernovae-driven winds originating from SNII explosions associated with ongoing star formation (see \citealt{Pillepich2018MNRAS.473.4077P} for more details). In general, these outflows are slower and do not escape to distances as large as black hole-driven outflows~\citep{nelson2019b}.

\subsection{Initial conditions and model variations}

When specifying the initial conditions for the gas, in addition to small density and velocity perturbations required to realize the chosen cosmological constraints, TNG must also specify the initial conditions for the magnetic fields\textcolor{red}{, which are given in terms of the strength and direction of the initial magnetic fields,} which can be very different for the different production mechanisms, i.e. astrophysical and primordial~\citep{garaldi2021}. All TNG simulations to date have been run with the same configuration of the initial magnetic field, which is assumed to be a constant volume-filling field (\textcolor{red}{which is an approximation} for a primordial, i.e. inflational magnetogenesis, \textcolor{red}{and which is the common practice in such simulations.} In this work we mainly use the TNG100 box with the initial strength $B_0=10^{-14}$~cG (comoving Gauss). At $z=127$ this corresponds to a \textcolor{red}{physical} field strength of $B_0 = 0.16$~nG.

In addition, a large number of `TNG model variation' simulations have been run, each changing a single parameter value or model choice to assess its importance in the fiducial TNG galaxy formation model~\citep{Pillepich2018MNRAS.473.4077P}. These simulate smaller $25~\text{cMpc/h}$ volumes, each realized at three resolution levels equivalent to those available for TNG100 itself. In this work, we primarily use \textcolor{red}{four simulations which vary the initial magnetic field strength, adopting $B_0=10^{-25}~\text{cG}$, $10^{-16}~\text{cG}$, $10^{-16}~\text{cG}$ (the fiducial choice of the TNG100 flagship run) and $10^{-12}~\text{cG}$}. We also inspect the outcome of three other runs that, compared to the TNG fiducial model, include no SMBH feedback, no SMBH kinetic feedback, and no feedback of any type, respectively.

\subsection{Analysis methodology}
\label{sec:analysis-methodology}

We analyze these simulations from $0\le z \le 6$ and measure the density of free electrons and magnetic field strength, avoiding $z > 6$ where these quantities become uncertain during the epoch of reionization~\citep{Heinrich:2016ojb,Aghanim:2018eyx}. The electron number density is calculated from the helium and hydrogen number densities and their ionization states. We use the spectral synthesis code \textsc{Cloudy V17.01}~\citep{Ferland2017RMxAA..53..385F} to calculate the ion fractions of hydrogen and helium for gas exposed to a redshift-dependent UV background~\citep{Haardt2012ApJ...746..125H}. We neglect the contribution of ionization states from elements heavier than He, as well as molecular gas phases.

\begin{figure*}
    \centering
    \includegraphics[width=1\textwidth]{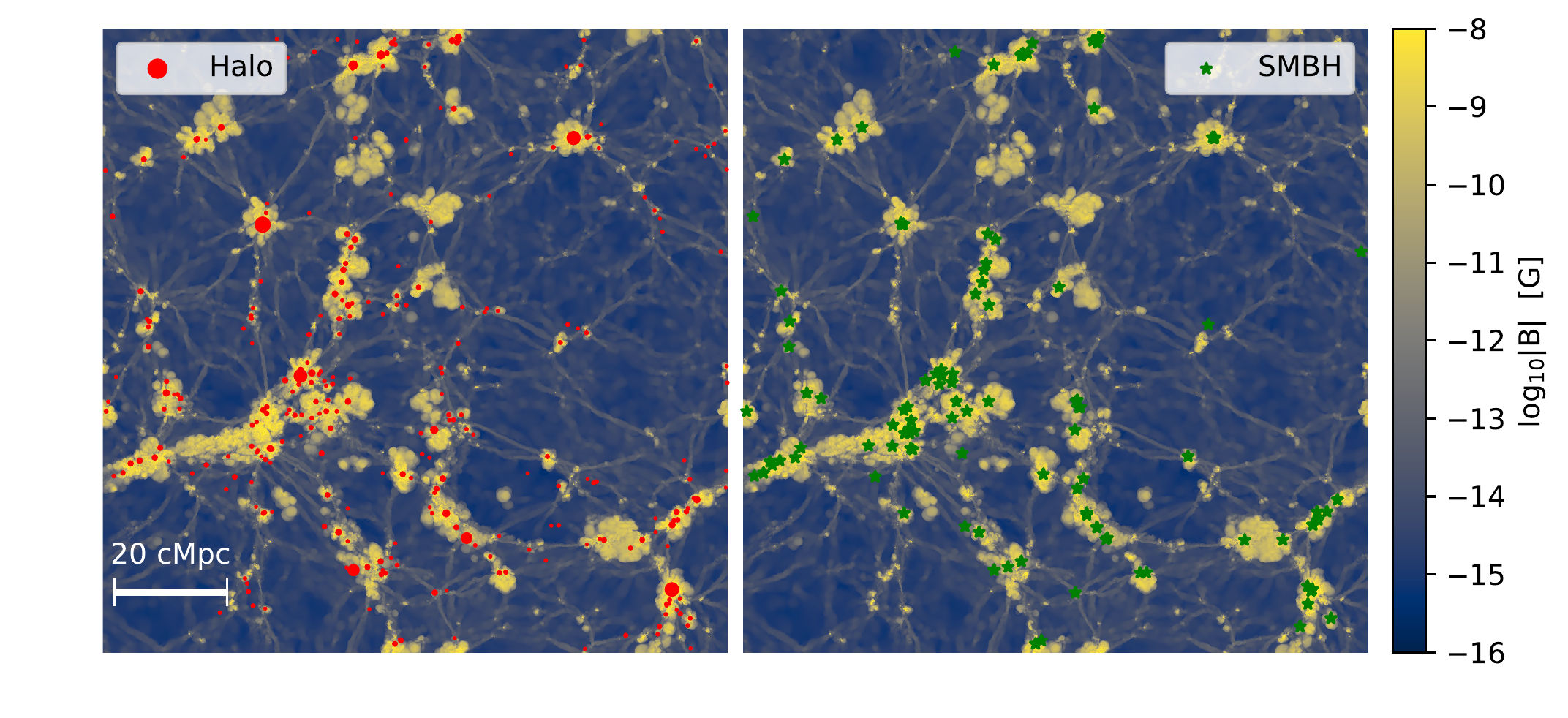}
    \caption{The magnetic field distribution for the same slice as in Fig.~\ref{fig:Bslice}, from the TNG100 simulation. Within $\pm 2.5$~Mpc of this slice we indicate all halos with mass $10^{11.5}M_{\odot}$ and above (red circles; left panel, radii corresponding to $1.5\,\text{R}_{200}$) and all black holes (stars; right panel) which have injected significant amount of energy back to the gas (with $E_{\rm low} > 10^{58.5}~\rm{ erg}$, see discussion in the text). Both are strongly correlated with the presence of large-scale magnetized bubbles.}
    \label{fig:agnhalos}
\end{figure*}

The magnetic field and electron number density are smoothed onto a grid with $(20~\text{ckpc})^3$ voxels.\footnote{We use the publicly available pysph-viewer code~\citep{pysph} for this deposition.} One thousand lines of sight (LOS) are generated for each available snapshot with random orientation within the simulated volume, \textcolor{red}{and we use these throughout as our fiducial set of sightlines}. 
In Fig.~\ref{fig:Bslice} we show an example of the magnitude of the magnetic field, metallicity, and electron number density image of a given region together with a potential LOS. The regions of magnetic field enhancement extend beyond gravitationally collapsed structures (i.e. dark matter halos), c.f. also Fig.~\ref{fig:agnhalos}.
They also extend, in some cases, to substantially larger distances, due to the ejection of magnetic fields in supernovae and black hole-driven galactic outflows~\citep{nelson2019b}.

\begin{figure*}
    \centering
    \includegraphics[width=0.48\textwidth]{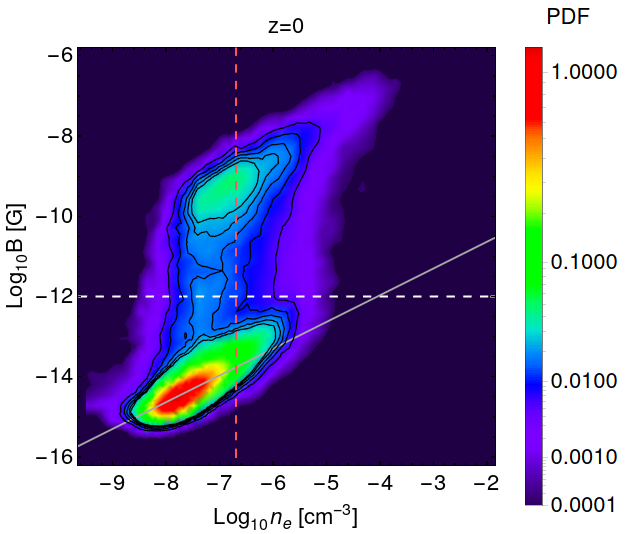}~\includegraphics[width=0.48\textwidth]{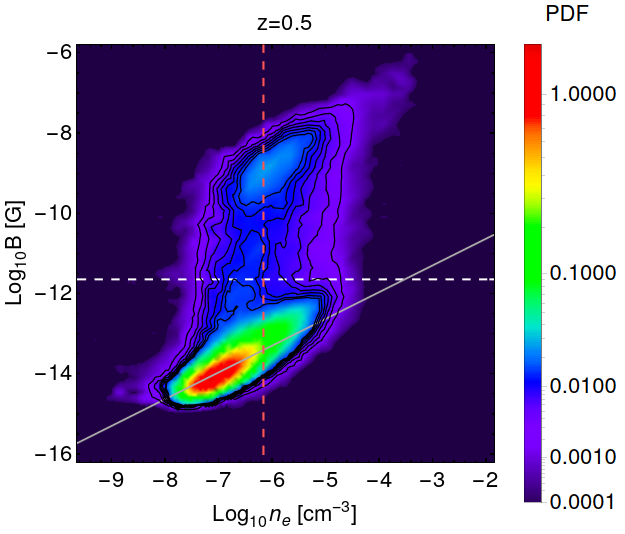} \\
    \includegraphics[width=0.48\textwidth]{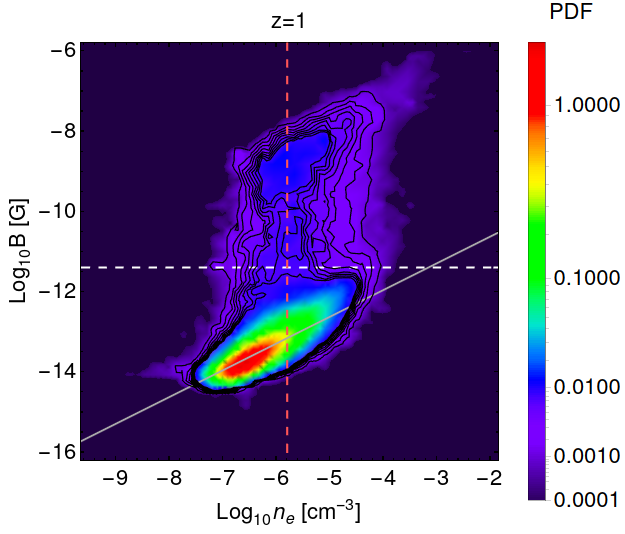}~\includegraphics[width=0.48\textwidth]{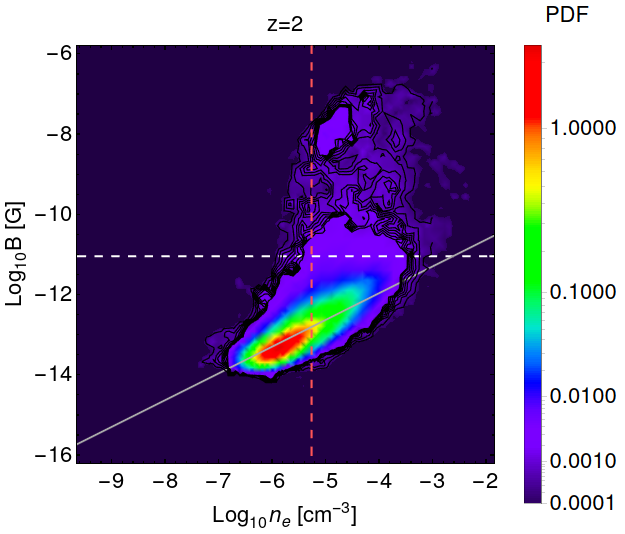}
    \caption{\textcolor{red}{Distribution of the magnetic field strength and electron number density in the TNG100 simulation using data along 200 random lines of sight with $(20\text{ ckpc})^3$ voxels in the box (see Sec.~\ref{sec:analysis-methodology})}, where the seed field is $B_0 = 10^{-14}$~cG, at redshifts $z=0$, $0.5$, $1$ and $2$. \textcolor{red}{The color indicates the occupied volume fraction}. The \textcolor{red}{dashed white} line corresponds to the comoving magnetic field value $10^{-12}$~cG that we use as the smallest value of outflow-generated magnetic fields in this work. The \textcolor{red}{red dashed} line represents the average electron number density at a given redshift. \textcolor{red}{The gray dashed line shows a power law $B\propto n_e^{2/3}$ that represents adiabatic evolution (see e.g.~\citealt{Durrer:2013pga}).} At fixed electron number density, two distinct branches of magnetic field strength are apparent, corresponding to weak and strong components, respectively. Results for other redshifts and simulation boxes are given in Appendices~\ref{app:25Mpc},~\ref{app:100Mpc}, and~\ref{app:300Mpc}.}
    \label{fig:ne_B}
\end{figure*}

\section{Statistical properties of simulated magnetic fields}
\label{sec:igm}

We start by examining the three \textcolor{red}{$25~\text{cMpc/h}$} variation boxes with different initial magnetic field seed conditions. In particular, we aim to determine to what extent the predicted IGMF depends on the value of the initial seed. In Fig.~\ref{fig:B3} (left panel) we show the mass-weighted distribution of the magnetic field for three different choices of the initial field. These global distributions across all gas cells contain two clear peaks -- a low-$B$ peak that has its center at the value of the initial field, and a strong-$B$ peak, that \textcolor{red}{are very similar for all four} values of the initial conditions.

It is important to keep in mind that the number of gas cells with a given value of the field does not represent the fraction of volume occupied by such a field. As the simulation is spatially adaptive and maintains a constant mass resolution, gas cells accumulate inside high-density regions. We therefore also show the volume-weighted distribution of magnetic field strength in Fig.~\ref{fig:B3} (right panel). Both cases show the same picture: that the high-$B$ component of the distribution is \textcolor{red}{weakly sensitive} to the magnitude of the initial field. \textcolor{red}{To characterize difference in strong-$B$ peaks numerically we calculated mass and volume weighted fractions within this peaks for different seed fields, see Table~\ref{tab:filling-fractions}. We see that while the seed field value changes by $13$ orders of magnitude, all fractions change by less then a factor $1.3$.}
\textcolor{red}{This is due to the rapid amplification processes which effectively erase knowledge of the primordial seed fields \citep[see also][]{pakmor14} explored in this work}.
Note that these distributions include not only galaxies and clusters but also larger volumes potentially affected by outflows, and the surrounding intergalactic medium, as we discuss in more details below.

\begin{figure*}
    \centering
    \includegraphics[width=0.48\textwidth]{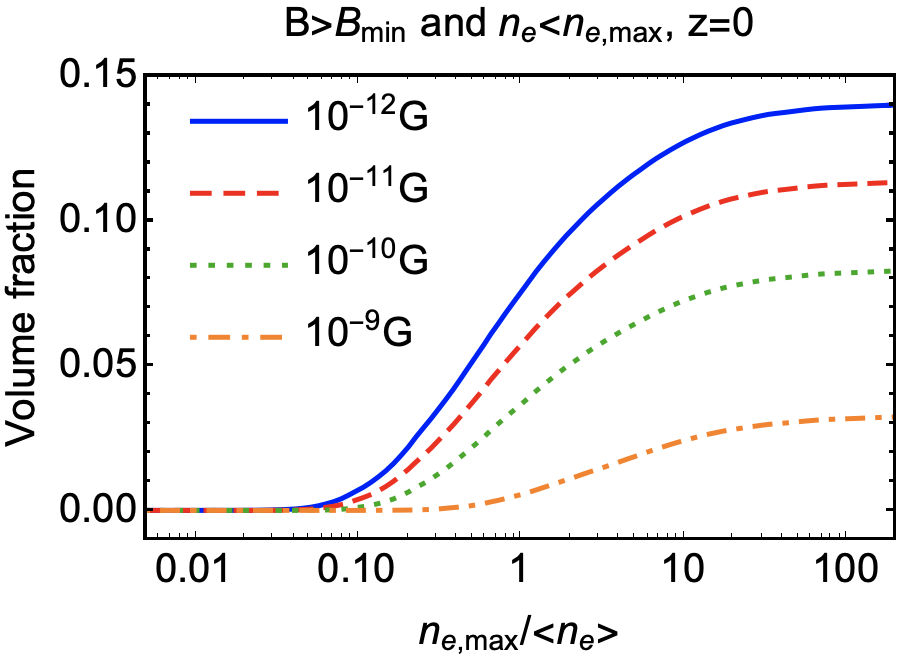}~\includegraphics[width=0.48\textwidth]{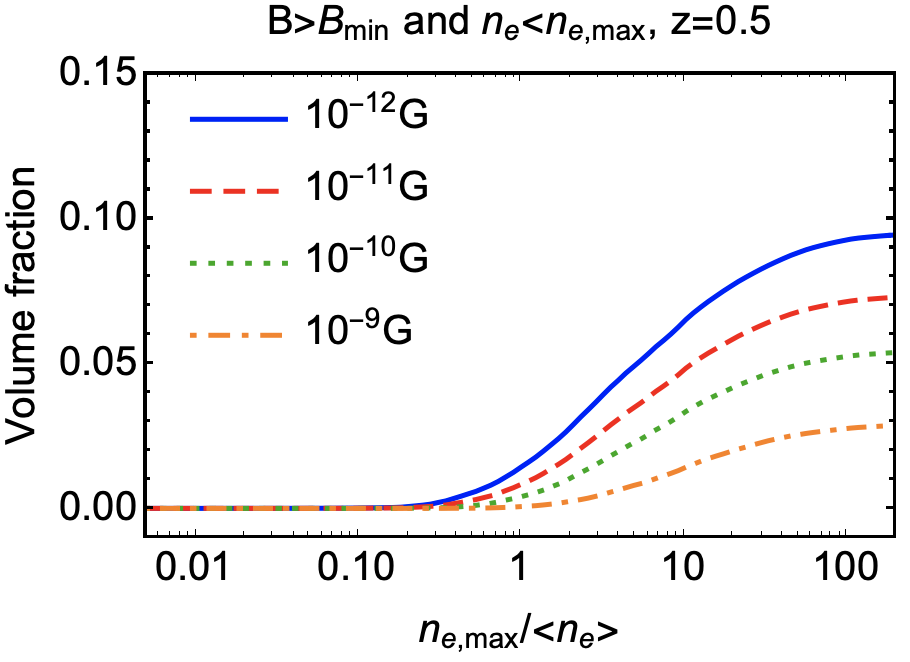}
    \caption{Volume fractions of the regions where the magnetic field is larger than $B_{\min}$ (legend) and the electron number density is smaller than $n_{e,\max}$ (x-axis) for $z=0$ (left panel) and $z=0.5$ (right panel). We here show TNG100 with a seed field of $B_0 = 10^{-14}$~cG \textcolor{red}{using data along 200 random lines of sight with $(20\text{ ckpc})^3$ voxels in the box (see Sec.~\ref{sec:analysis-methodology})}. Comparing these two redshifts, a substantial growth of the volumes occupied by high magnetic field strengths is evident, implying that physical processes within the last 5\,Gyr have had a strong impact. Results for other redshifts and simulation boxes are given in Appendices~\ref{app:25Mpc},~\ref{app:100Mpc}, and~\ref{app:300Mpc}.}
    \label{fig:volume-fraction}
\end{figure*}

\begin{figure*}
    \centering
    \includegraphics[width=0.48\textwidth]{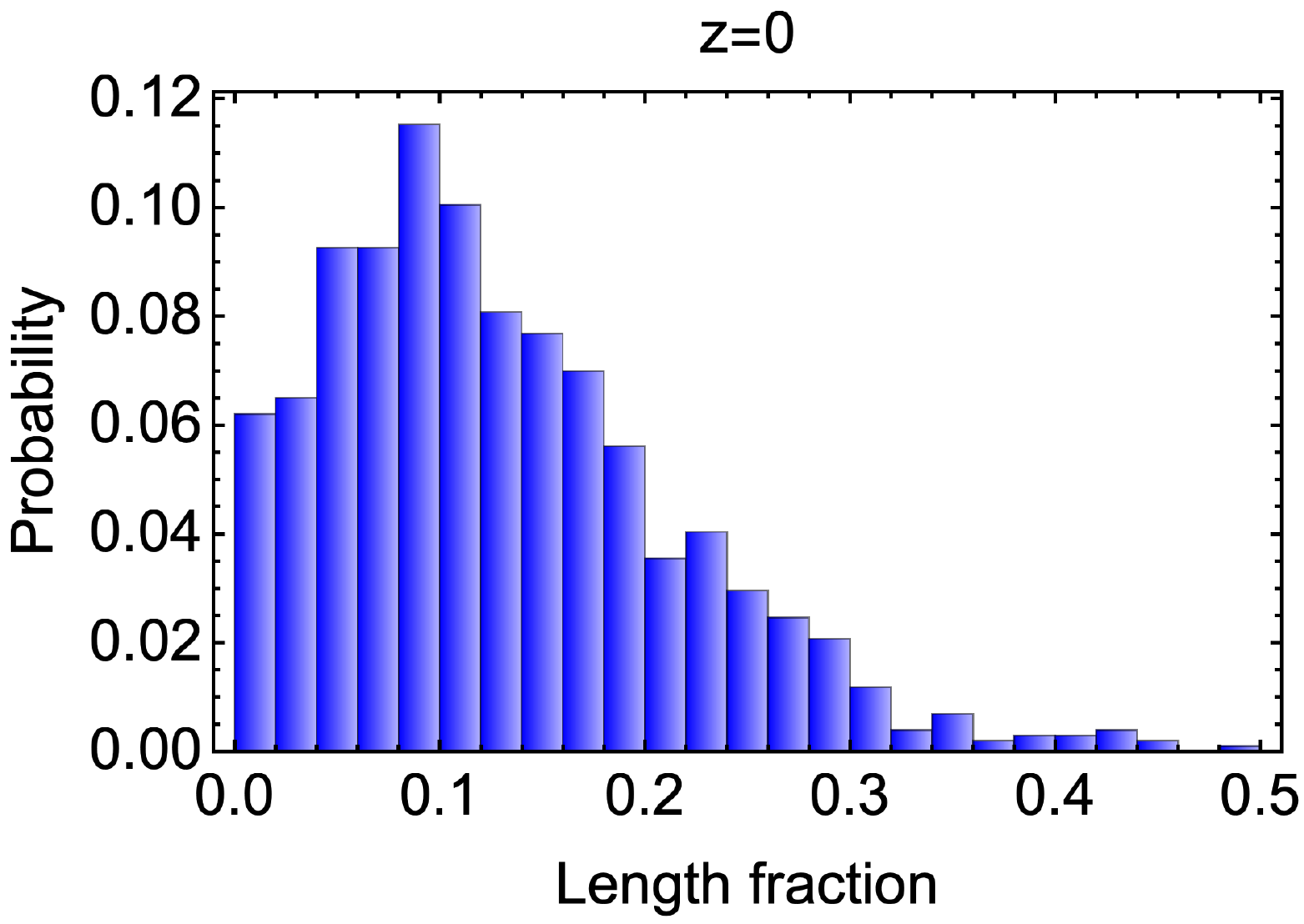}
    \includegraphics[width=0.48\textwidth]{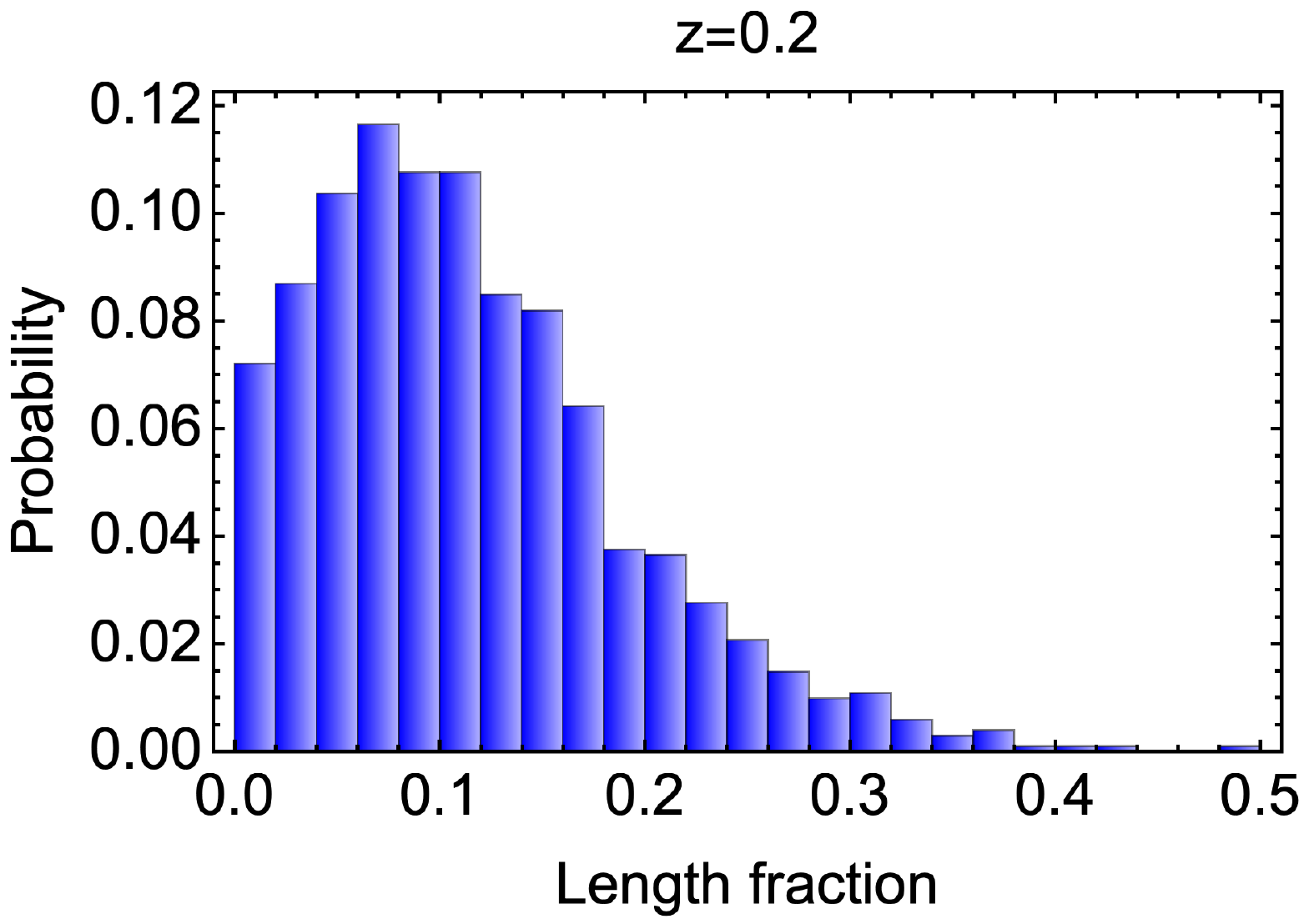}
    \includegraphics[width=0.48\textwidth]{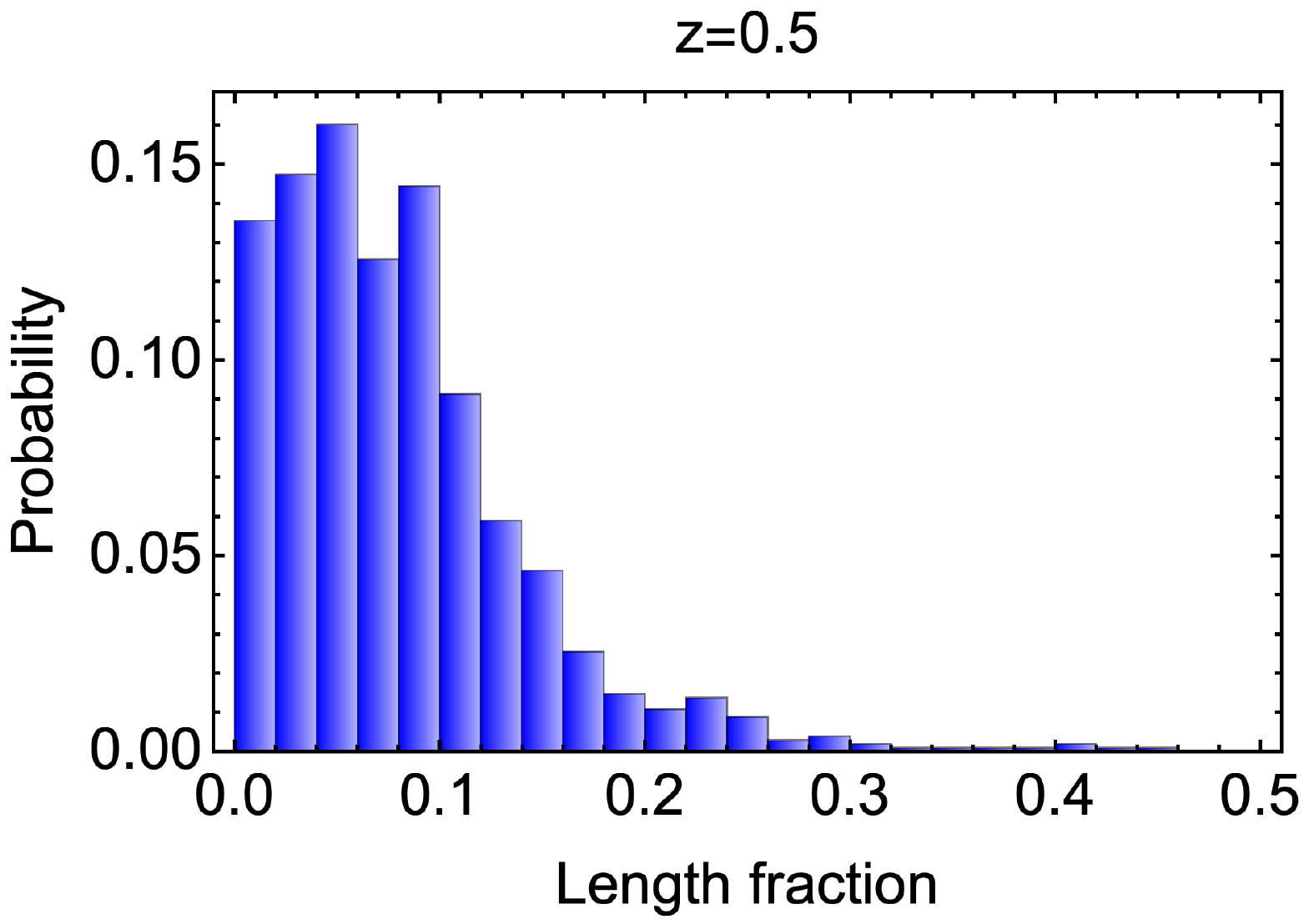}
    \includegraphics[width=0.48\textwidth]{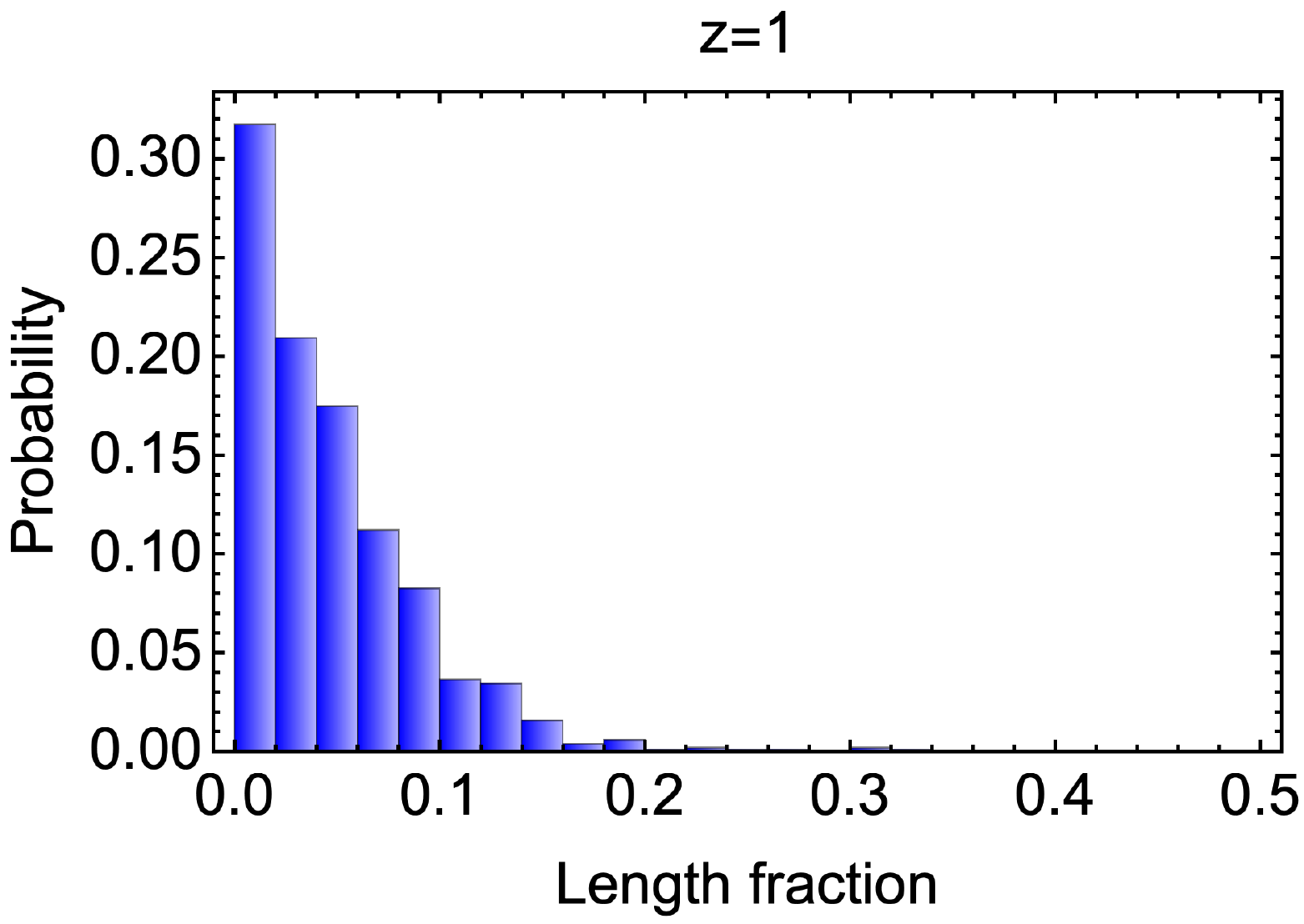}
    \caption{\textcolor{red}{Probability to find a fraction of length along the line of sight with magnetic field larger than $10^{-12}$ comoving Gauss.}
    We show results for TNG100 at redshifts $z=0$, $0.2$, $0.5$, and $1$, creating 1000 random lines of sight at each redshift. While there is negligible `strong magnetic field path length' at high redshifts, by $z=0$ roughly half of all sightlines intersect such strong magnetic fields along more than 10 percent of their length. Results for other redshifts and simulation boxes are given in Appendices~\ref{app:100Mpc} and~\ref{app:300Mpc}.}
    \label{fig:length-fraction}
\end{figure*}

\begin{figure*}
    \centering
    \includegraphics[width=0.48\textwidth]{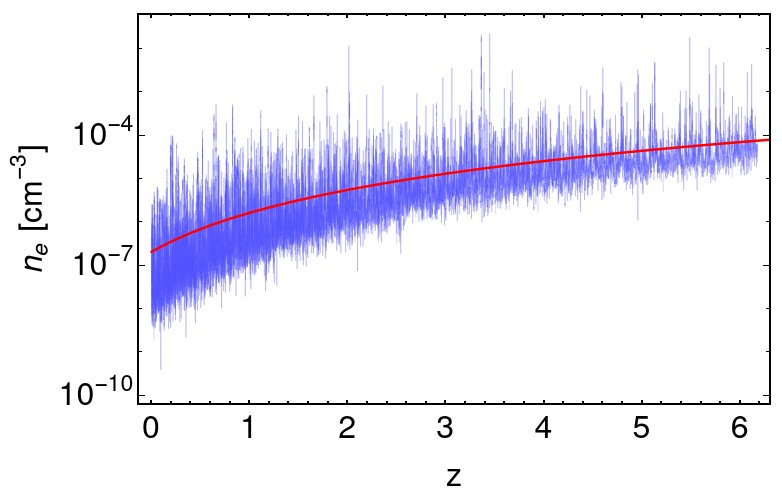}~\includegraphics[width=0.48\textwidth]{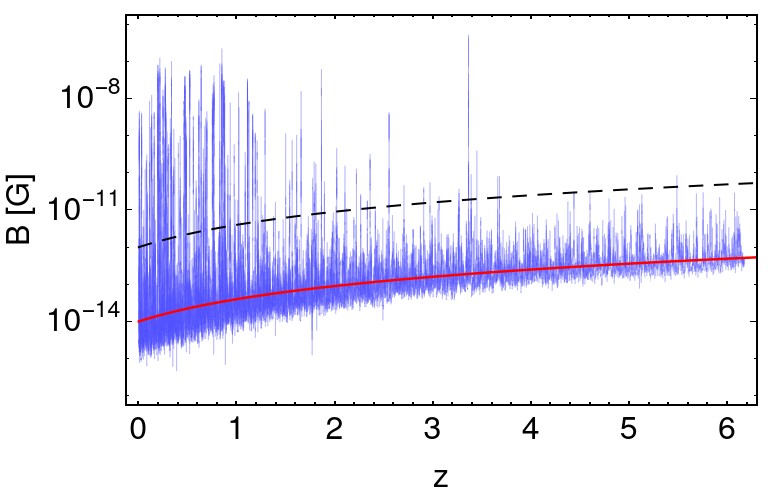}
    \caption{An example along a single continuous line of sight through the TNG100 simulation: electron number density (left panel) and magnetic field strength (right panel). The red line on the left panel shows the average electron number density in the Universe. The red line on the right panel corresponds to the seed value of the magnetic field, while the black line shows $B = 10^{-12}$~cG.}
    \label{fig:continuousLOS}
\end{figure*}

To study the \textcolor{red}{large-value component} of the magnetic field more quantitatively we analyze the main simulation box TNG100 with a seed magnetic field $10^{-14}$~cG. Let us consider its distribution versus electron number density at different redshifts, shown in Fig.~\ref{fig:ne_B} (see also~\citealt{marinacci18}). At low redshift $z \lesssim 2$ we see two distinct branches, corresponding to weak and strong magnetic fields, respectively. In both branches, the value of the magnetic field is correlated with the electron number density. Even in the regions with small values of electron number density $\sim 10^{-8}\text{ cm}^{-3}$ (one-tenth of the average electron number density today), the magnetic field can be many orders of magnitude stronger than its average value (close to the value of the initial field). This strong-$B$ (or ``over-magnetized'') branch in Fig.~\ref{fig:ne_B} corresponds to the primordial seed independent strong-$B$ peak in Fig.~\ref{fig:B3}. In Appendices~\ref{app:25Mpc}, \ref{app:100Mpc}, \ref{app:300Mpc} we present plots similar to Fig.~\ref{fig:ne_B} that illustrate that two branches also exist for $z<2$ for different initial conditions and different box sizes, while for $z>2$ there is generically only one branch in the $B$-$n_e$ plot, due to the time needed for exponential amplification via a small-scale turbulent dynamo. \textcolor{red}{This amplification process is faster at higher numerical resolution, enabling the magnetic fields to reach their quasi-saturated values earlier, although the level of this final saturation is relatively unaffected by resolution~\citep{pakmor2017}. In such a relatively small volume, there are also no high-mass halos at early times ($z \gtrsim 3$), such that large outflow-driven bubbles have not yet formed. Either larger volumes, i.e. probing the environments of the largest overdensities, or higher resolution would therefore if anything enhance the importance of these structures.}

Our goal is to understand the impact of the outflow-generated large-$B$ component of the magnetic field distribution on the propagation of light through the Universe. Specifically, what is the probability for a photon to occupy a large-$B$ region on its way towards an observer? Fig.~\ref{fig:volume-fraction} therefore shows the volume fraction of regions with large magnetic field values, excluding the regions where the electron concentration is larger than some given value (in this way we can, in particular, exclude the inner parts of the collapsed structures like galaxies).\footnote{For similar figures for different seed field values and box sizes see Figs.~\ref{fig:25Mpc_vfrac},~\ref{fig:volume-fraction3}, and~\ref{fig:volume-fraction2}.}
At $z=0$ in regions with $n_e < 100 \langle n_e \rangle$ the magnetic field is stronger than $10^{-12}$~cG in $\sim 14\%$ of the volume, while it is stronger than $10^{-10}$~cG in $\sim 8\%$ of the volume. We note that the value of the initial field is $10^{-14}$~cG. Moving only to $z=0.5$, we see that strong-$B$ regions occupy half as much volume, indicating that strong-$B$ regions are substantially enhanced by late time processes. 

Alternatively, we can measure the fractional length, for a given line of sight, which intersects a strong magnetic field. Using our sample of 1000 random sightlines we show in Fig.~\ref{fig:length-fraction} the distribution of length fractions having magnetic field strength $B\geq 10^{-12}$~cG at four different redshifts.\footnote{For similar figures for different seed field values and box sizes see Fig.~\ref{fig:length-fraction3}.} At $z=0$ more than half of all sightlines intersect such strong magnetic fields along more than 10\% of their path length. The peak of the distribution shifts to smaller fractional path lengths towards higher redshift. At $z\geq1$, sightlines intersect magnetic fields of this strength only rarely.

For further applications, it is also interesting to consider longer lines of sight that do not fit into one $\sim$\,100 cMpc snapshot. Using the TNG simulation volume we constructed continuous lines of sight from $z=0$ to $z=6$ following the procedure described in~\cite{Bondarenko:2020moh}. To construct the magnetic field along a continuous line of sight we take $B(z)$ from additional random sightlines within the same snapshot, and assign it to any missing pathlength (between simulation snapshots) of the continuous LOS, rescaling as $B\propto (1+z)^2$. An example of both electron number density and magnetic field strength along a single continuous sightline is shown in Fig.~\ref{fig:continuousLOS}.

\subsection{Connection between over-magnetized `bubbles' and galaxy outflows}

Visual inspection of Fig.~\ref{fig:Bslice} reveals the strong connection between over-magnetized bubbles (i.e. the regions where magnetic fields are orders of magnitude larger than in the average regions with the same values of $n_e$, see Fig.~\ref{fig:ne_B} and discussion), \textcolor{red}{metallicity}, and the structures seen in the electron number density. This is consistent with the behavior seen in the TNG galaxy formation model where strong outflows can escape from galaxies and break out into the intergalactic medium, carrying mass, heavy elements, and magnetic energy density along the way~\citep{nelson2019b}. \textcolor{red}{This behavior is also consistent with results for galactic and near-galactic magnetic fields discussed in~\cite{Butsky:2015pya,2020MNRAS.498.3125P,2020MNRAS.495.4475M}.}

\begin{figure*}
    \centering
    \includegraphics[width=.8\textwidth]{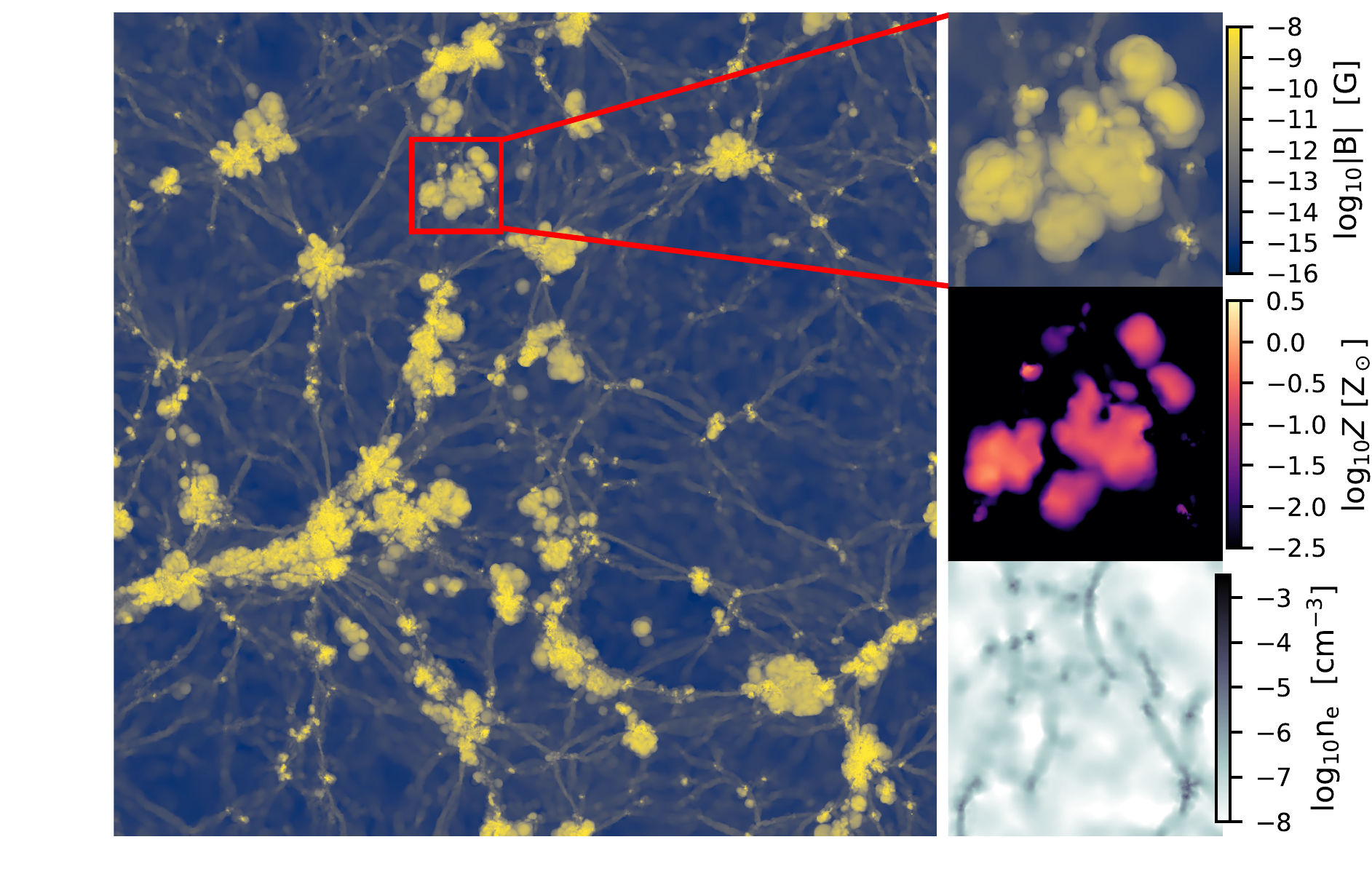}
    \caption{Zoom on a magnetized bubble region from TNG100 selected to have no supermassive black holes satisfying the energy threshold, $E_{\rm low} > 10^{58.5}~\rm{ erg}$, within 5 Mpc of the slice. Further, this region is not clearly associated with an overdensity in $n_e$, implying that some large-scale magnetized bubbles can arise from the combined action of many lower mass galaxies, hosting less effective black holes and/or supernovae-driven outflows.}
    \label{fig:zoom}
\end{figure*}

To explore the physical connection between large-scale over-magnetized `bubbles' and galactic-scale feedback processes, Figure \ref{fig:agnhalos} shows the magnetic field strength in thin ($20$~kpc) slices of TNG100 at $z=0$. In the left panel, we mark all dark matter halos with total mass $> 10^{11.5}$\,M$_{\odot}$ with red circles, where the marker size denotes $1.5\,\rm{R}_{\rm 200}$, where $\rm{R}_{\rm 200}$ is the virial radius of a halo. In the right panel, we instead mark all supermassive black holes which have injected a significant amount of energy in the low-accretion state, $E_{\rm low} > 10^{58.5}~\rm{ erg}$, with stars. We see that most magnetized bubbles extending to $\gtrsim$\,Mpc scales are directly associated with massive halos and/or supermassive black holes near their center.

\begin{figure*}
    \centering
    \includegraphics[width=0.9\textwidth]{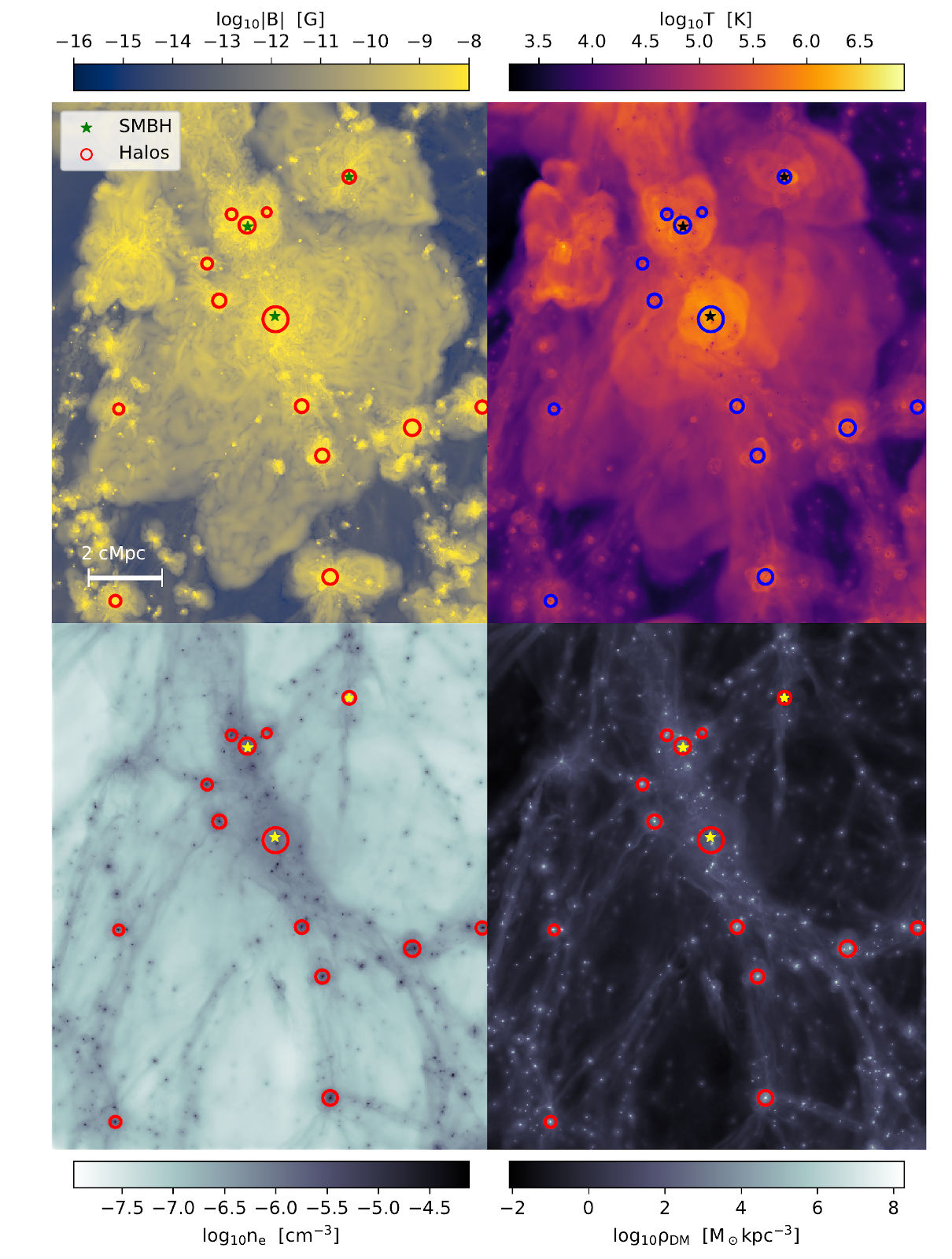}
    \caption{View of a rich, high-overdensity structure from TNG100 which hosts a large-scale, $\sim 10$\,Mpc, outflow-driven bubble \textcolor{red}{that corresponds to the same large over-magnetized bubble from Fig.~\ref{fig:zoom}}. The top-left panel shows magnetic field strength, next to it the gas temperature. The \textcolor{red}{bottom-left} panel shows electron number density and the in the \textcolor{red}{bottom-right} we have dark matter density in the region. The three most active supermassive black holes, marked by stars, are largely responsible for the extent of over-magnetized regions.}
    \label{fig:blend}
\end{figure*}

This is, however, not always the case. Figure \ref{fig:zoom} zooms into a crowded region of the same slice, within which an over-magnetized bubble is evident. The three panels on the right show magnetic field strength, gas metallicity, and electron number density. There is no clear association between cosmic web filaments visible in $n_e$ and regions of strong $|B|$. We have specifically selected this region as having no supermassive black holes which satisfy our energy criterion. Their absence implies that either the SMBH sourcing this bubble is outside the $\pm 2.5$ Mpc vicinity of the slice or that this large bubble may be a collective effect of galactic-scale winds produced by ongoing supernovae explosions in lower mass, star-forming galaxies, possibly in combination with the effect of a number of smaller black holes.

\begin{figure*}
    \centering
    \includegraphics[width=0.9\textwidth]{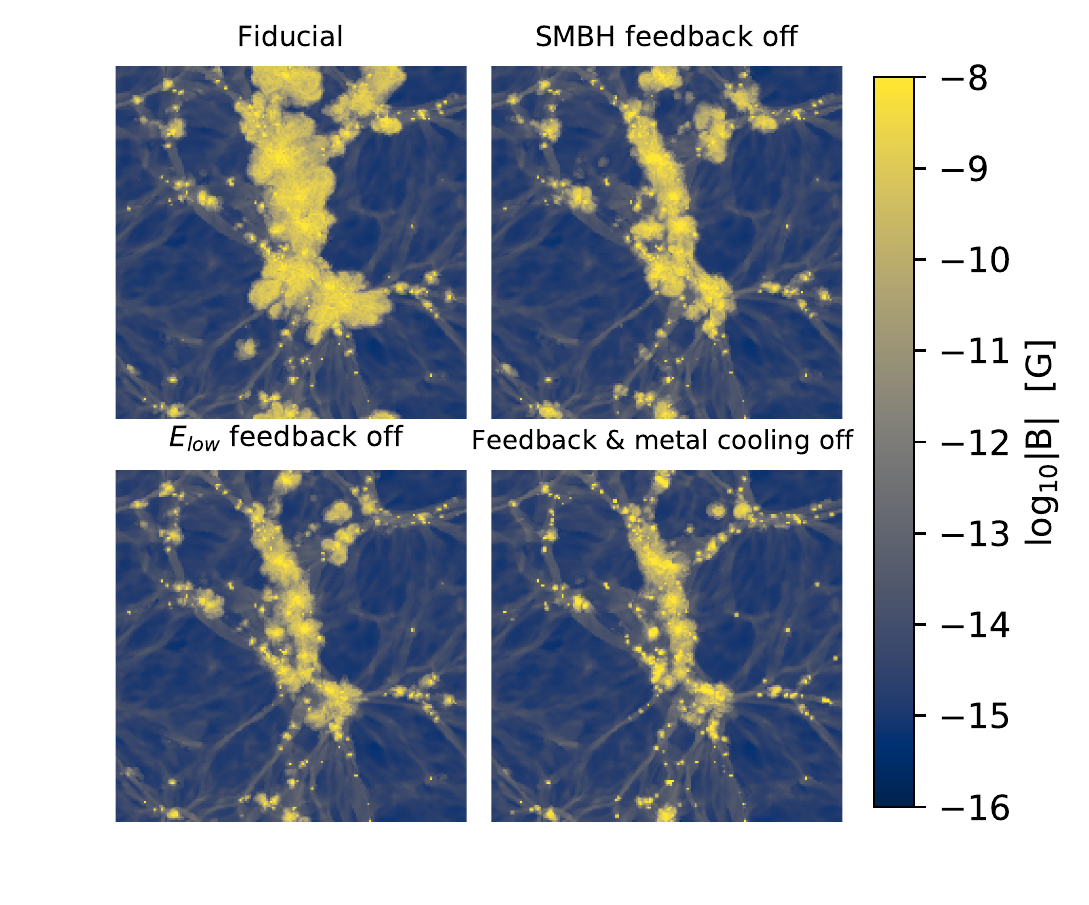}
    \caption{A $(25\text{ Mpc})^2 \times 1$~Mpc slice of the magnetic field from four \textcolor{red}{25 cMpc/h} simulations at $z=0$. We contrast the fiducial TNG model (upper left) to a variation with no black hole feedback of any kind (upper right), only SMBH low-state kinetic winds disabled (lower left), and no feedback from black holes nor supernovae at all (lower right). In each variation panel there are fewer and smaller large-scale magnetic bubbles with respect to the fiducial model, with the efficient SMBH kinetic wind mode playing a dominant role.}
    \label{fig:noBH}
\end{figure*}

To better understand the origin of these bubbles, Fig.~\ref{fig:blend} shows the magnetic field, gas temperature, dark matter, and electron number densities in a region extending 12~Mpc $\times$ 14.1~Mpc $\times$ 15~Mpc that contains the same large over-magnetized bubble from the Fig.~\ref{fig:zoom}.
In this volume, we see a clear filament of large-scale structure. We again mark the massive halos and energetic black holes. Particularly clear around the SMBHs in the center and upper right are signatures of a collimated, episodic outflow made up of successive gas shells, forming a butterfly-like morphology, which results from the SMBH kinetic wind feedback. Here the ejection of magnetic fields into these bubbles is directly caused by active SMBHs. 

However, stellar feedback can also contribute, most notably through core-collapse supernovae (SN) explosions. These also produce galactic-scale outflows, albeit in lower mass galaxies. To isolate these two mechanisms, we turn to additional simulations of the `TNG model variation' suite. In particular, Fig. \ref{fig:noBH} compares the fiducial model (upper left) to models with no SMBH kinetic wind feedback (lower left), no SMBH feedback of any kind (upper right), and no feedback whatsoever from either black holes or supernovae (lower right). In all cases, the identical \textcolor{red}{25 cMpc/h} volume is shown. The regions present in the fiducial run, but missing when the kinetic winds are disabled, are due to the low-accretion feedback mode of the SMBHs. The overall extent of the central filamentary region is such an example. Furthermore, regions present in the runs without black hole feedback, but missing in the `no feedback' case, are due to supernovae -- some of the smaller bubbles towards the upper right being prime examples. In general, bubbles inflated by SN rather than SMBHs tend to be smaller, have lower temperatures, and lower 
expansion velocities.
Overall, we see that both SMBHs and SN contribute to extended regions of high magnetic energy, with the supernovae playing a sub-dominant role.

\section{Axion conversion probability}
\label{sec:axion}

In this section, we apply our findings to calculate the probability of photon-axion conversion during propagation through the IGM. This conversion occurs in the presence of an external magnetic field $B$ due to the interaction term
\begin{equation}
    \mathcal{L}_{a \gamma} = 
    \frac{g_a}{4} a F_{\mu \nu} \Tilde{F}_{\mu \nu} = g_a a (\partial_0 A_i) \cdot B_i,
\end{equation}
where $g_a$ is the ALP constant (with units of inverse energy), $B_i$ are components of the magnetic field, and $A_i$ are spatial components of $A_{\mu}$ in the gauge $A_0 = 0$. For ALPs with energy $E_a$ in an external magnetic field this interaction effectively gives mass-mixing between the ALP and the photon,
\begin{equation}
    \mathcal{L}_{a \gamma} = g_a E_a B_i a A_i = g_a E_a B_T a |\bm{A}|,
\end{equation}
where $B_T = |B|\cos\theta$ and $\theta$ is an angle between vector $\bm{A}$ and the magnetic field. This leads to oscillations between axions and photons~\citep{1988PhRvD..37.1237R}. The strength of the mixing between axions and photons is proportional to the magnetic field as well as the energy of the axion.

The conversion probability also depends on the axion mass $m_a$ and the effective photon mass in the medium. For soft enough photons propagating through IGM the effective photon mass is given by (see e.g.~\citealt{Mirizzi:2009nq})
\begin{equation}
    m_{A}(n_e) = \sqrt{\frac{4\pi \alpha_{\text{EM}} n_e}{m_e}},
    \label{eq:mA_ne}
\end{equation}
\textcolor{red}{where $m_e$ is the electron mass, $\alpha_{\text{EM}}$ is the fine-structure constant and $n_e$ is a free electron number density}. It should be noted that for gamma-ray photons this description of the effect on the effective properties of photons may not be sufficient (see e.g.~\citealt{Dobrynina:2014qba}). However, in this paper we deal with less energetic photons and therefore we can use Eq.~\eqref{eq:mA_ne} throughout.

In the case when the effective photon mass is equal to the ALP mass, $m_a = m_{A}(n_e)$, the conversion becomes resonant and the conversion probability significantly increases. The conversion probability for this case is~\citep{Mirizzi:2009nq}
\begin{equation}
    P_{a \to A} = 1 - p, \qquad 
    p = \exp\left(-\frac{\pi E_a B_T^2 g_a^2}{ m_{a}^2} R\right),
\end{equation}
where $R = \left|d \log m_A^2/d\ell \right|^{-1}_{\ell = \ell_{res}}$, $E_a$ is the axion energy, $B_{T}$ is the component of the magnetic field orthogonal to the line of sight (direction of axion propagation), $\ell$ is the distance along the line of sight, and the derivative for the $R$ factor is calculated at the point of the resonance. The case $p\ll 1$ is called the adiabatic limit and the conversion probability is close to one, while the case $p \approx 1$ is called the non-adiabatic limit and the conversion probability is given by
\begin{equation}
    P_{a \to A}^{\text{non-adiab}} = \frac{\pi E_a B_T^2 g_a^2}{ m_{a}^2} R.
\end{equation}

Let us assume that the total conversion probability $P_{\text{tot}}$ during axion propagation along the LOS is much smaller than one. In this case the total conversion probability is given by
\begin{equation}
    P_{\text{tot}} \approx \sum_i P_{a \to A,i}^{\text{non-adiab}} = \frac{\pi E_a g_a^2}{ m_{a}^2} \sum_i (1+ z_i) B_{T,i}^2 R_{i},
    \label{eq:ptot}
\end{equation}
where $E_a$ is the axion energy at $z=0$, and the sum is taken for resonances along the line of sight: $z_i$ is the redshift and $B_{T,i}$ is the component of the magnetic field orthogonal to the line of sight for a given resonance. 

\begin{figure}
    \centering
    \includegraphics[width=\columnwidth]{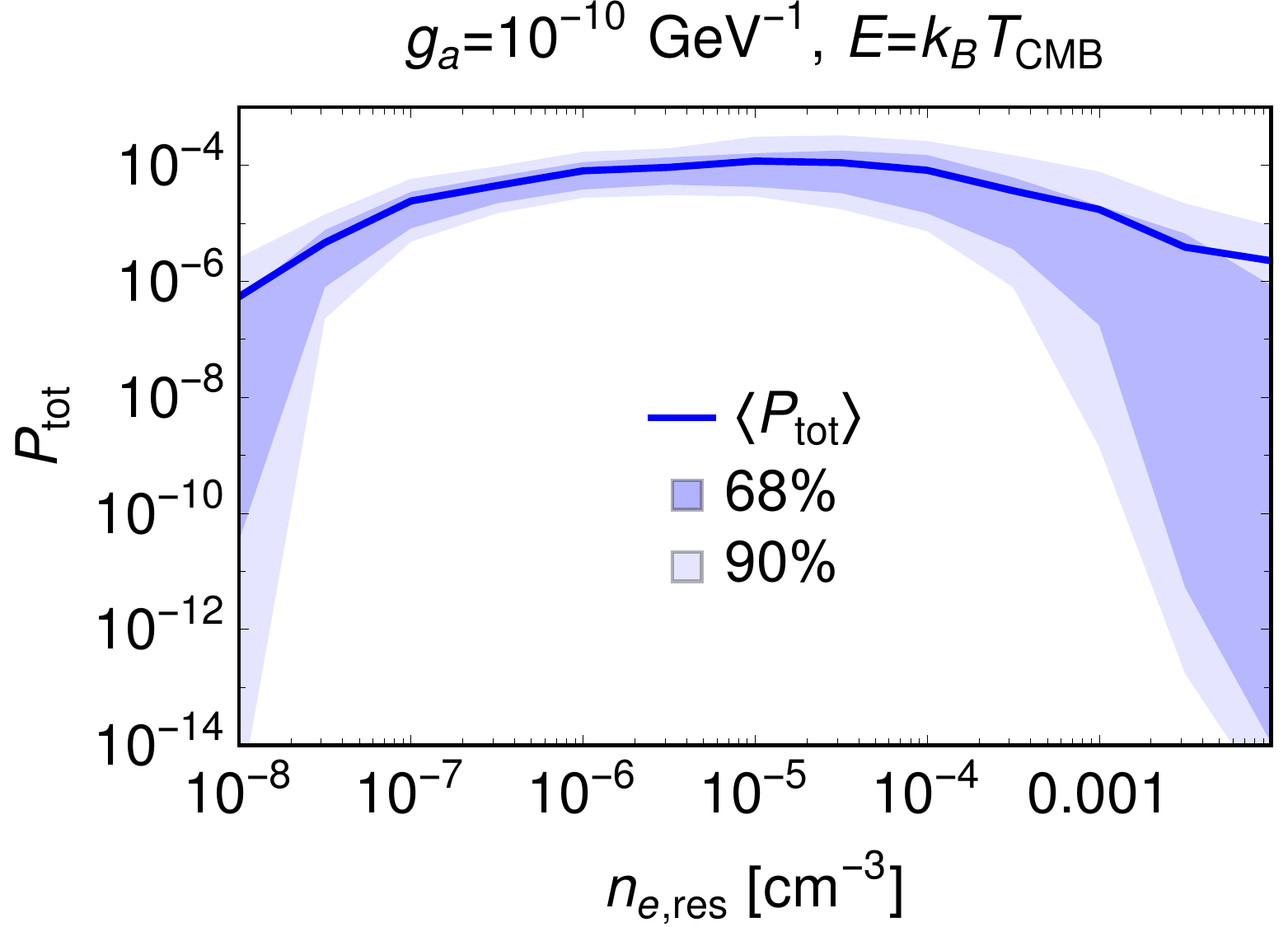}
    \caption{The conversion probability versus resonant electron number density for ALP energy $E=k_B T_{\text{CMB}}$, $1/g_a = 10^{10}$~GeV calculated over 500 random continuous lines of sight from TNG100 up to redshift $z=6$. The blue line represents the average values, and the shaded regions correspond to $68\%$ and $90\%$ of conversion probabilities.}
    \label{fig:ptot}
\end{figure}

\subsection{Probability of axion conversion along random sightlines}

As the mixing strength between axions and photons is proportional to the magnetic field strength squared, the probability of the axion-photon conversion is dominated by the contribution of the strong-$B$ component of the IGMF distribution. As we have seen above, this part of the distribution is to a large extent universal -- its dependence on the value of the initial seed field is negligible (at least within the IllustrisTNG framework and within the class of initial conditions used in this paper). Further study of the dependence on initial conditions from a wider class (e.g. those of~\citealt{Vazza:2017qge}) and with different models of baryonic physics will be considered in future work.

We proceed with our analysis based on the TNG simulations with initial magnetic field strength $10^{-14}$~cG. We take into account only the contribution to conversion from the simulation pixels with the value of magnetic field $B\geq 10^{-12}$~cG (see the black line in Figs.~\ref{fig:ne_B} and~\ref{fig:continuousLOS}). 
\textcolor{red}{As we discussed in Section~\ref{sec:igm}, for such magnetic field values, the distribution of magnetic field only slightly depends on the properties of the primordial seed field for seed fields $10^{-14}$~cG and below. As a result, this threshold on magnetic field strength (in comoving units) returns a somewhat conservative contribution to the axion conversion probability from the magnetic bubbles.}

Using 500 simulated continuous lines of sight (as described in Section~\ref{sec:analysis-methodology}), we calculate the average conversion probability and its distribution between redshifts $z=0$ and $z=6$. The result as a function of resonant electron number density is shown in Fig.~\ref{fig:ptot}. We see that conversion probability is maximal near $n_{e,\text{res}} = 10^{-5}\text{ cm}^{-3}$. The scatter grows for large and small values of $n_{e,\text{res}}$ because of the small amount of resonances along the line of sight. In the region of resonant electron number densities below $10^{-8}\text{ cm}^{-3}$ or above $10^{-2}\text{ cm}^{-3}$ the resonant condition occurs rarely (see Fig.~\ref{fig:continuousLOS}). We therefore derive  below the constraints on the axion-photon coupling for the axion mass range $4\cdot 10^{-15}\text{ eV} < m_a < 4\cdot 10^{-12}$~eV.

\subsection{Constraints from CMB distortions}

In this section, we consider the effect where a CMB photon converts into an axion. The probability of this conversion is proportional to the photon energy, and its occurrence induces distortions in the CMB spectrum. The strength of the effect depends on the axion coupling $g_a$, so we can constrain the axion model from CMB spectrum measurements obtained by COBE/FIRAS~\citep{Fixsen:1996nj}, which determined the CMB spectrum in the frequency range from 68 to 637 GHz with a precision of up to $\Delta B_E/B_E \approx 10^{-4}$, where $B_E$ is a measured spectral radiance and $\Delta B_E$ is its uncertainty.

Resonant axion-photon conversion modifies the CMB spectrum as
\begin{align}
    B_E (E) = B^{\text{CMB}}_{E}(E) [1 -  \langle P_{\text{tot}}(E)\rangle ],
    \label{Btot}
\end{align}
where $B^{\text{CMB}}_{E}$ is the spectral radiance of the initial CMB spectrum and $\langle P_{\text{tot}}(E)\rangle$ is the average conversion probability. We see that conversion produces an energy-dependent modification of the Planck spectrum. We estimate the exclusion region of the COBE/FIRAS measurement by a simple condition that follows from~\eqref{Btot},
\begin{equation}
    \frac{\Delta B_E}{B_E} = \langle P_{\text{tot}}(k_B T_{\text{CMB}})\rangle < 10^{-4},
    \label{eq:FIRAS_simple_estimate}
\end{equation}
where $T_{\text{CMB}} =  2.7260(13)$~K~\citep{Fixsen:2009ug}.

\begin{figure}
    \centering
    \includegraphics[width=\columnwidth]{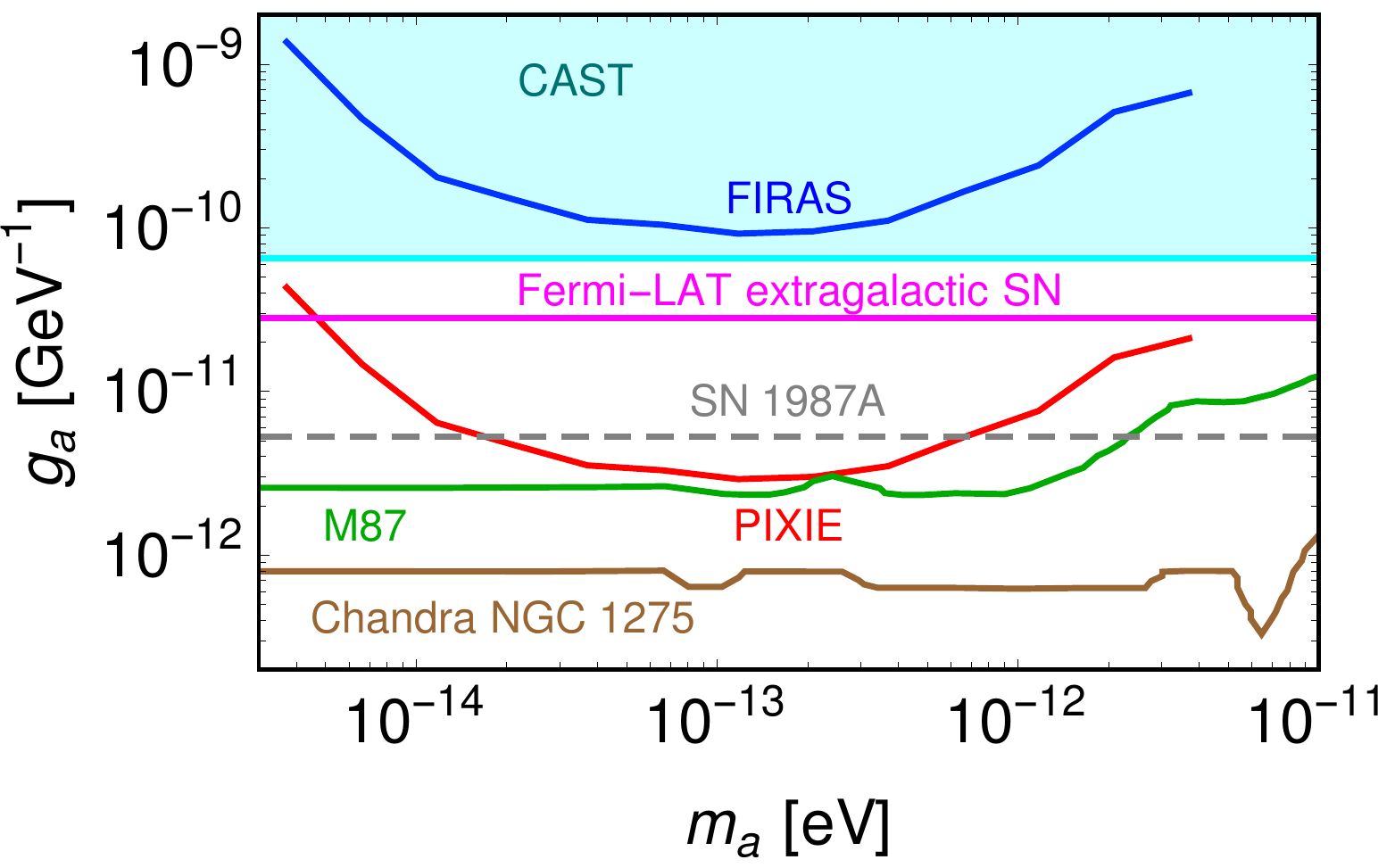}
    \caption[]{Constraints on the axion-photon coupling constant, $g_a$, as a function of axion mass, $m_a$, from FIRAS/COBE (blue line) and a projection of the future sensitivity of the PIXIE experiment (red line). We also present here other relevant constraints in this mass range of the axion: CAST~\protect\citep{Anastassopoulos:2017ftl}, Fermi-LAT extragalactic SN~\protect\citep{Meyer:2020vzy}, M87~\protect\citep{Marsh:2017yvc}, Chandra NGC 1255~\protect\citep{Reynolds:2019uqt}.}
    \label{fig:fa}
\end{figure}

The result for the exclusion is shown in Fig.~\ref{fig:fa}, where we also add an estimation of the sensitivity of the future CMB distortion experiment PIXIE using the same condition~\eqref{eq:FIRAS_simple_estimate} but taking an expected sensitivity for the PIXIE experiment of $\Delta B_E/B_E < 10^{-7}$~\citep{Chluba:2019nxa,Chluba:2019kpb}. As we can see, the CMB-based constraints are not competitive with other existing constraints. The reason for this is clear -- the probability of axion-photon conversion is proportional to energy (see Eq.~\eqref{eq:ptot}). Therefore, much stronger constraints can be obtained from sources of photons with higher energy, e.g. X-ray or, especially, gamma-rays, see e.g.~\cite{Montanino:2017ara,Reynolds:2019uqt}, where however non-resonant conversion is discussed. X-ray and gamma-ray sources are not all-sky, but individual point sources, and a study of the effect of the IGM on their spectra requires a different methodology that is beyond the scope of this paper. We expect however that our results can be used for such an analysis in the future.

\section{Summary, Discussion and Conclusions}
\label{sec:conclusion}

In this paper, we have quantified the effects of galaxy evolution processes on the magnetization of the intergalactic medium (IGM). To this end, we have used several simulations from the IllustrisTNG~\citep{nelson18,marinacci18} suite which all include treatment of ideal magnetohydrodynamics coupled to a state-of-the-art model for galaxy formation physics and feedback. In these calculations we always assume a homogeneous initial seed field, which is then amplified throughout the process of collapse and structure formation.

As demonstrated in previous analyses \textcolor{red}{(see e.g.~\citealt{pakmor2017})}, strong magnetic fields are produced inside galaxies due to small-scale MHD dynamos. In this paper, we have shown that such strongly amplified magnetic fields can be distributed to larger volumes due to galactic feedback, in particular feedback from supermassive black holes \textcolor{red}{(see also~\citealt{nelson2019b})}. These large-scale bubbles produced by outflows from galaxies and clusters develop particularly at redshifts $z<2$, and contain magnetic fields that are several orders of magnitude stronger than in the unaffected regions of the IGM with the same electron density.

As a result, similarly to the magnetic fields inside galaxies, these fields are to a certain extent invariant with respect to the assumed initial conditions for magnetic fields (see Fig.~\ref{fig:B3} and its discussion in the text). We show that these over-magnetized bubbles with $|B| \ge 10^{-12}$~cG, enhanced metallicity, and with clear outflowing kinematic signatures, can be as large as tens of Mpc. Their existence, and extent, is directly related to feedback activity from supermassive black holes in the centers of massive galaxies. Supernovae explosions also produce similar albeit smaller magnetized bubbles around lower mass galaxies.

We study the volume filling fraction of these strong field regions and their distribution over random lines of sight. We find that a typical intergalactic line of sight at $z=0$ extends for \textcolor{red}{$\sim 10-15$\%} of its length within these regions \textcolor{red}{for seed magnetic fields lower than $10^{-14}$~cG}. 
This implies that strongly magnetized bubbles are important for the propagation of light and high-energy cosmic rays through the Universe.

We use these results from TNG to put bounds on the photon-axion conversion from spectral distortions of the CMB. The disappearance of CMB photons due to their resonant conversion into axions in the IGM introduces deviations from the black-body spectrum of the CMB. If the mass of the axion is in the range $4\cdot 10^{-15}\text{ eV} < m_a < 4\cdot 10^{-12}$~eV where many resonances happen along a typical LOS, this distortion can be above the limit of COBE/FIRAS. As the photon-axion coupling grows with energy, the bounds based on the CMB distortions are not competitive with other existing bounds obtained using much more energetic photons. 

In addition to CMB spectrum distortion, the conversion of CMB photons into axions produces additional anisotropy in both the temperature and polarization of the CMB. These bounds could be stronger than those obtained using the spectrum averaged over the whole sky. Indeed, the effect of the axion-photon conversion in the IGM is dominated by the contributions of the over-magnetized bubbles discussed in the paper. On one hand, this makes this effect to some extent independent of the unknown properties of the initial seed fields and therefore justifies the use of simulations. On the other hand, this means that anisotropies are introduced on the scales that are small for the CMB, i.e. the signal peaks at relatively high-$\ell$, where the power spectrum of CMB anisotropies is already suppressed. At the same time, the size of the bubbles is noticeably larger than that of dark matter halos themselves, so the resulting impact could be distinguished from e.g. the Sunyaev–Zeldovich effect. However, to perform such an analysis we cannot calculate the conversion probability only for infinitesimal sightlines, and must instead produce an anisotropy map for a part of the sky, which is a challenging task for future work.

Stronger constraints from photon-axion conversion in IGM could also leverage the energy dependence of the coupling constant in order to study the propagation of gamma-ray photons through the IGM (see e.g.~\citealt{Montanino:2017ara} for a study of the non-resonant conversion of gamma-rays in the IGM using ENZO simulations). This involves different classes of sources and will also be studied elsewhere.

\begin{figure}
    \centering
    \includegraphics[width=\columnwidth]{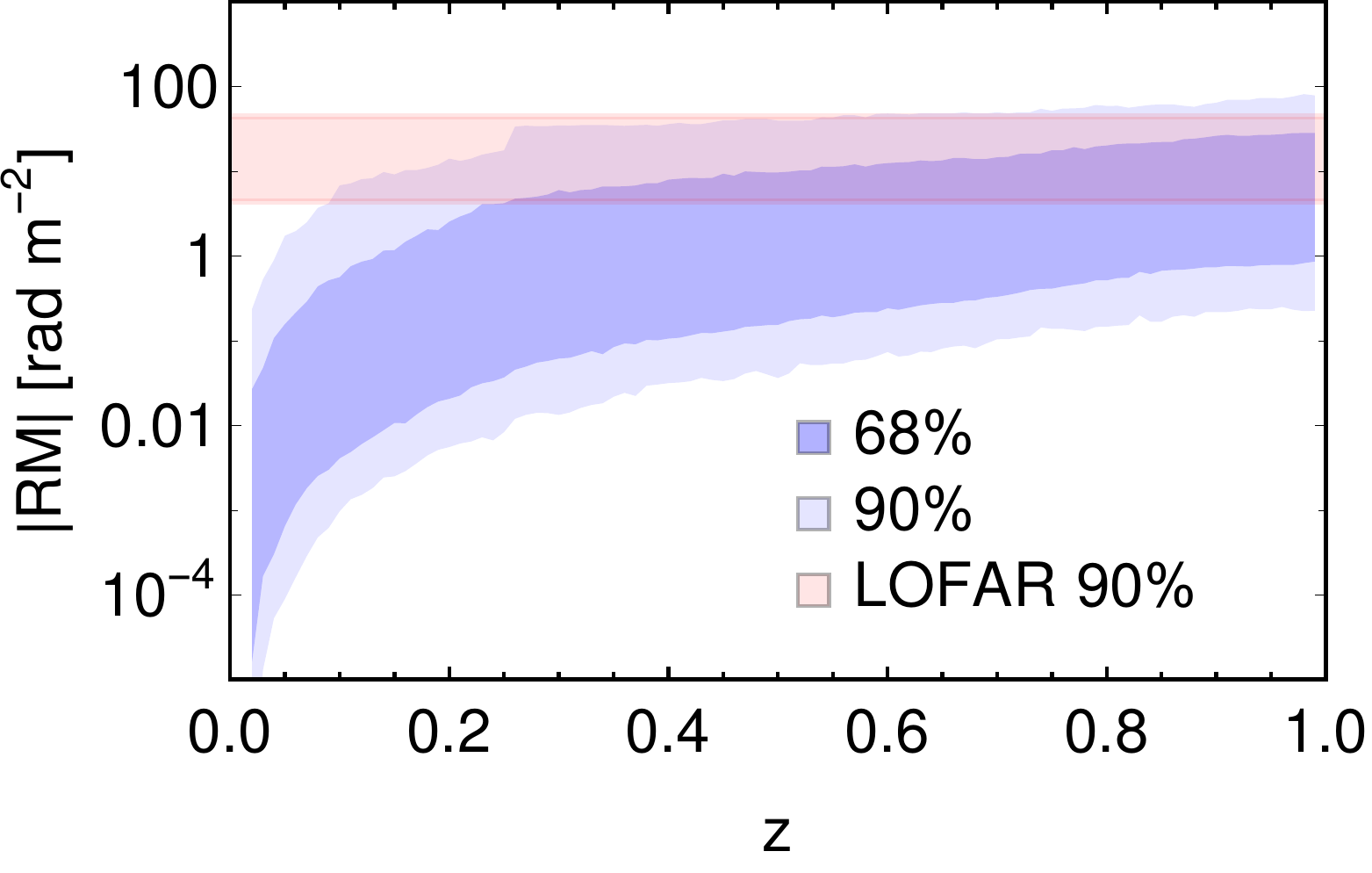}
    \caption[]{Distribution of the contribution to Faraday Rotation Measure from the magnetic fields in outflow-generated bubbles only (see text), for sources located at redshifts between $z=0$ and $1$ calculated using 1000 random lines-of-sight from the TNG100 simulation (see~\ref{sec:sim} for description). Blue (light blue) region contains 68\% (90\%) of all values of RM that we obtained for this mock lines-of-sight. The red region contains 90\% of values of RM of the observed radio sources presented in~\cite{OSullivan:2020pll}. }
    \label{fig:RM}
\end{figure}

Finally, the over-magnetized bubbles may contribute to measurements of the Faraday rotation measure (RM). For example, \cite{OSullivan:2020pll} studies 349 pairs of radio sources in order to separate the contribution to RM  from the IGM versus from the contribution of the local environments of the source and the observer (Milky Way), aiming to put a bound on the value of the primordial seed fields. However, the contribution of the over-magnetized bubbles encountered on the way from the source to the observer may be significant and could hinder constraints on the contribution of the ``true'' intergalactic magnetic fields that exist in regions unaffected by outflows.

To illustrate this, we show in Fig~\ref{fig:RM} the value of the Faraday Rotation measure for 1000 LOS as a function of the distance to the source (blue shaded bands). To disentangle the contribution of the bubbles, we choose the lines of sight such that they do not have magnetic field $B > 10^{-12}$~cG at the beginning and at the end of the LOS. We also exclude the contribution of voxels with electron density $n_e > 0.01$~cm$^{-3}$, to remove galaxies, and of voxels with magnetic fields $B < 10^{-12}$~cG, to include only the contribution of the bubbles. We also show the region that contains 90\% of values of RM of the radio sources observed by LOFAR presented in~\cite{OSullivan:2020pll} (red shaded band).

We see that for the sources located at $z \ge 0.2$ the contribution of outflow-driven bubbles to RM measurements can be significant. This contamination must therefore be taken into account when inferring intergalactic magnetic field values from RM measurements. On the other hand, a more detailed analysis could constrain the properties (i.e. size scales, abundance) of feedback-driven bubbles themselves, thus informing models of galaxy formation and feedback, particularly when radio source counterparts with measurable distances are identifiable.

We conclude by noting that the baryonic feedback models and physics employed in the IllustrisTNG simulations are necessarily simplified treatments. In the future, more sophisticated black hole feedback models, such as those modeling unresolved accretion disks, black hole spin, and relativistic jet production could produce different emergent outflows. Similarly, the TNG galaxy formation model neglects several physical processes, most notably cosmic rays, low-temperature cooling and chemistry below $10^4$\,K, and coupled radiation-hydrodynamical interactions, which could similarly impact the generation and propagation of the outflows driven by galaxies and their supermassive black holes.


\section*{Acknowledgements}

We thank Matthieu Schaller and Josef Pradler for the useful discussions.
AS is supported by the FWF Research Group grant FG1. AAG, KB and AB are supported by the European Research Council (ERC) Advanced Grant ``NuBSM'' (694896). The TNG simulations were carried out with compute time granted by the Gauss Centre for Supercomputing (GCS) under Large-Scale Projects GCS-DWAR and GCS-ILLU on the GCS share of the supercomputer Hazel Hen at the High Performance Computing Center Stuttgart (HLRS). Additional simulations and analysis were carried out on the Isaac cluster of the Max Planck Institute for Astronomy (MPIA), and the systems of the Max Planck Computing and Data Facility (MPCDF). 

\section*{DATA AVAILABILITY}

The data underlying this article is available on reasonable request.

\bibliography{refs}

\appendix

\section{Varying the initial seed field with 25 cMpc/h simulations}
\label{app:25Mpc}

An important aspect of our work is to determine dependencies on the initial conditions of the magnetic field seed. For this, we use the TNG model variations for \textcolor{red}{four different values of initial seeds}. Each contains $512^3$ gas and dark matter particles, within a box of $L \sim 25~\text{cMpc/h}$, making the resolution similar to that of TNG100.

\begin{figure*}
    \centering
    \includegraphics[width=0.41\textwidth]{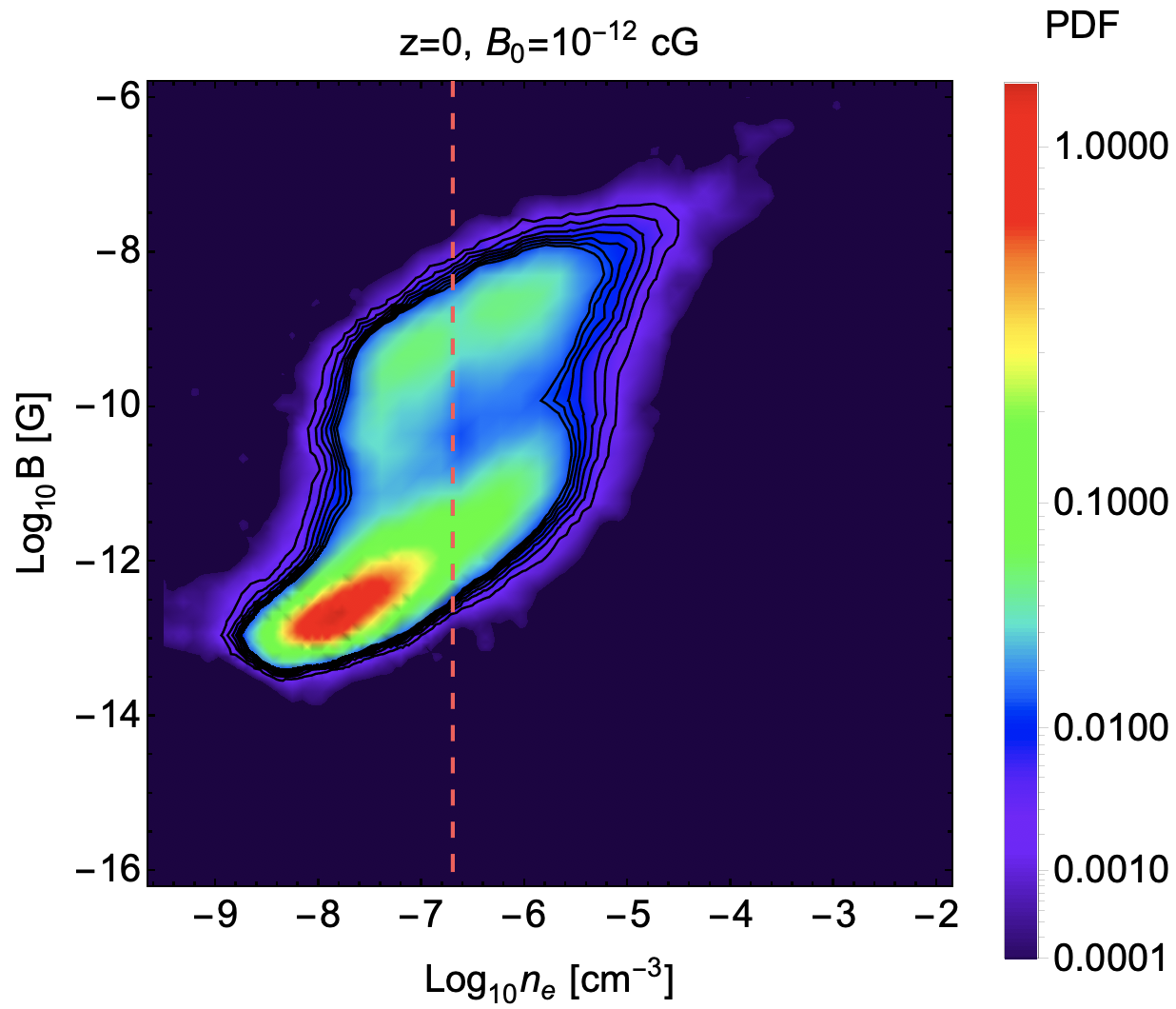}~\includegraphics[width=0.41\textwidth]{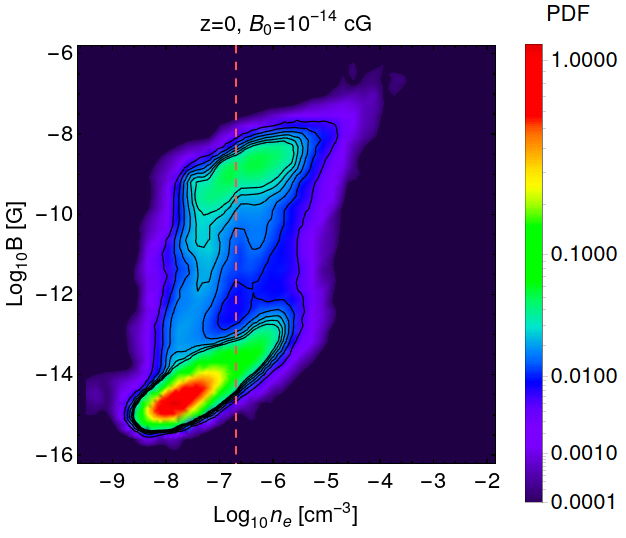} \\
    \includegraphics[width=0.41\textwidth]{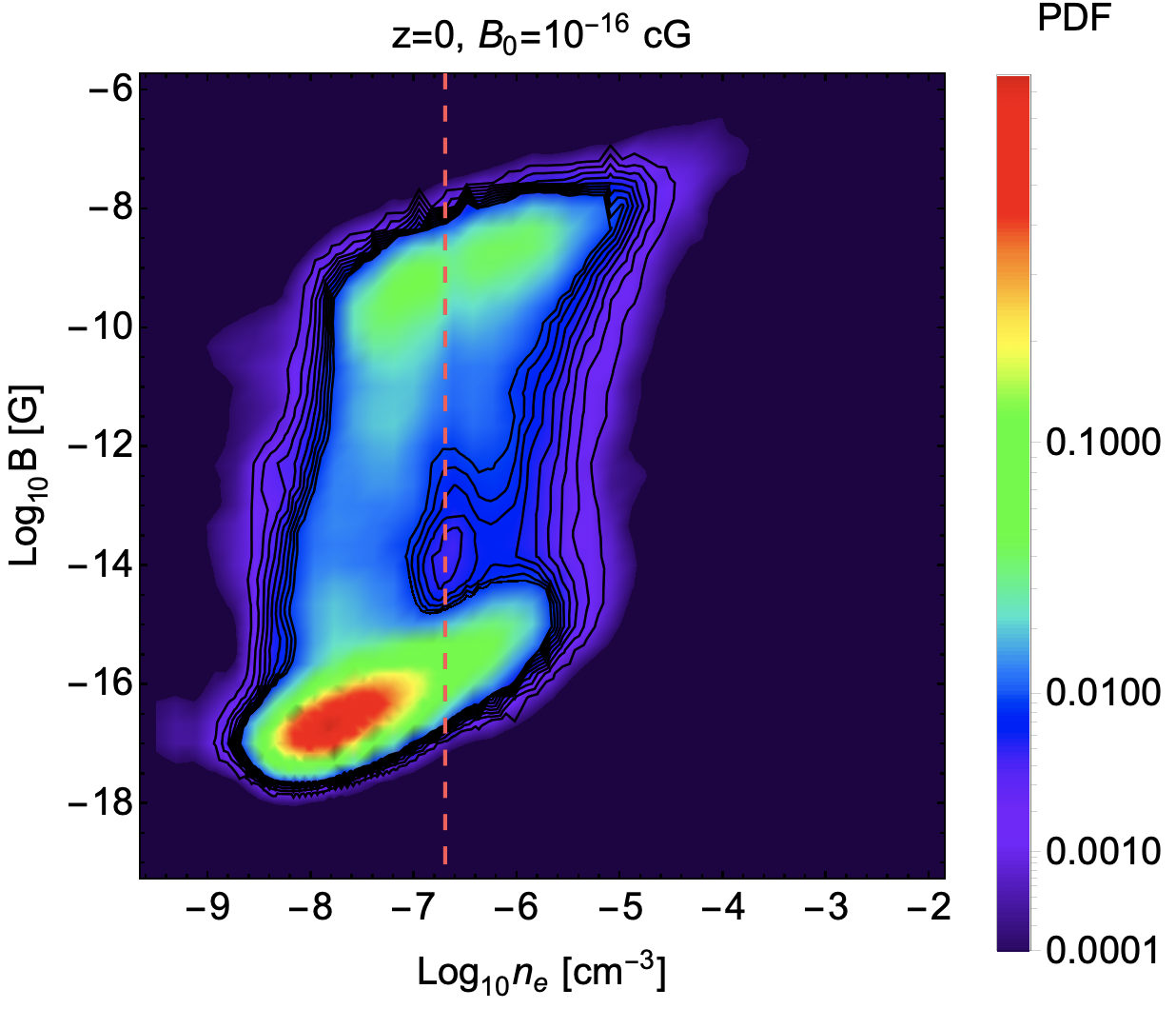}~\includegraphics[width=0.41\textwidth]{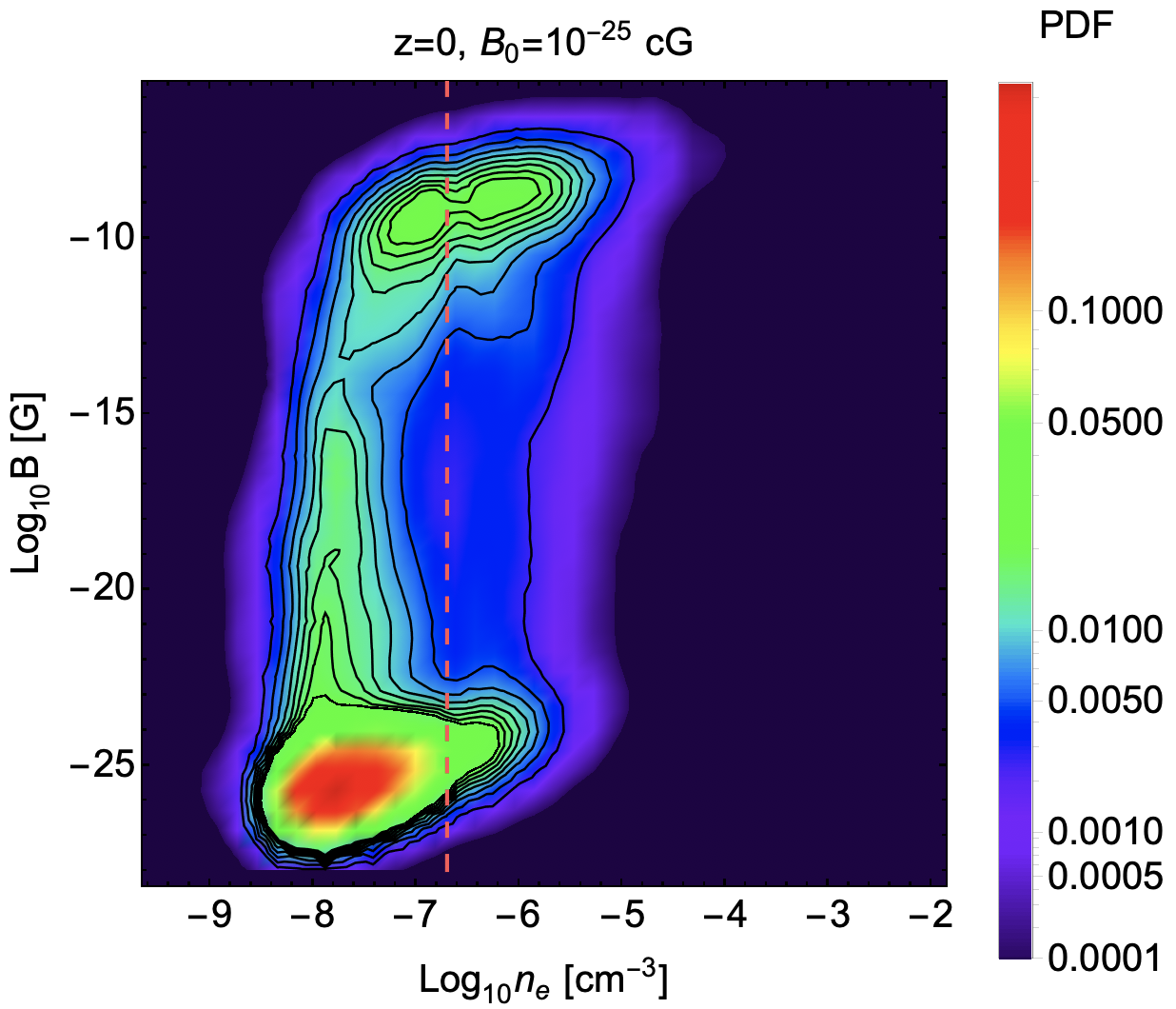}
    \caption{\textcolor{red}{Distribution of the magnetic field magnitude and electron number density in the fiducial 25 cMpc/h simulation} at redshift $z=0$ for different values of seed field: $10^{-12}$~cG, \textcolor{red}{$10^{-14}$~cG}, $10^{-16}$~cG, and $10^{-25}$~cG. The \textcolor{red}{red dashed} line represents the average electron number density.}
    \label{fig:25Mpc_B_ne}
\end{figure*}

\begin{figure*}
    \centering
    \includegraphics[width=0.37\textwidth]{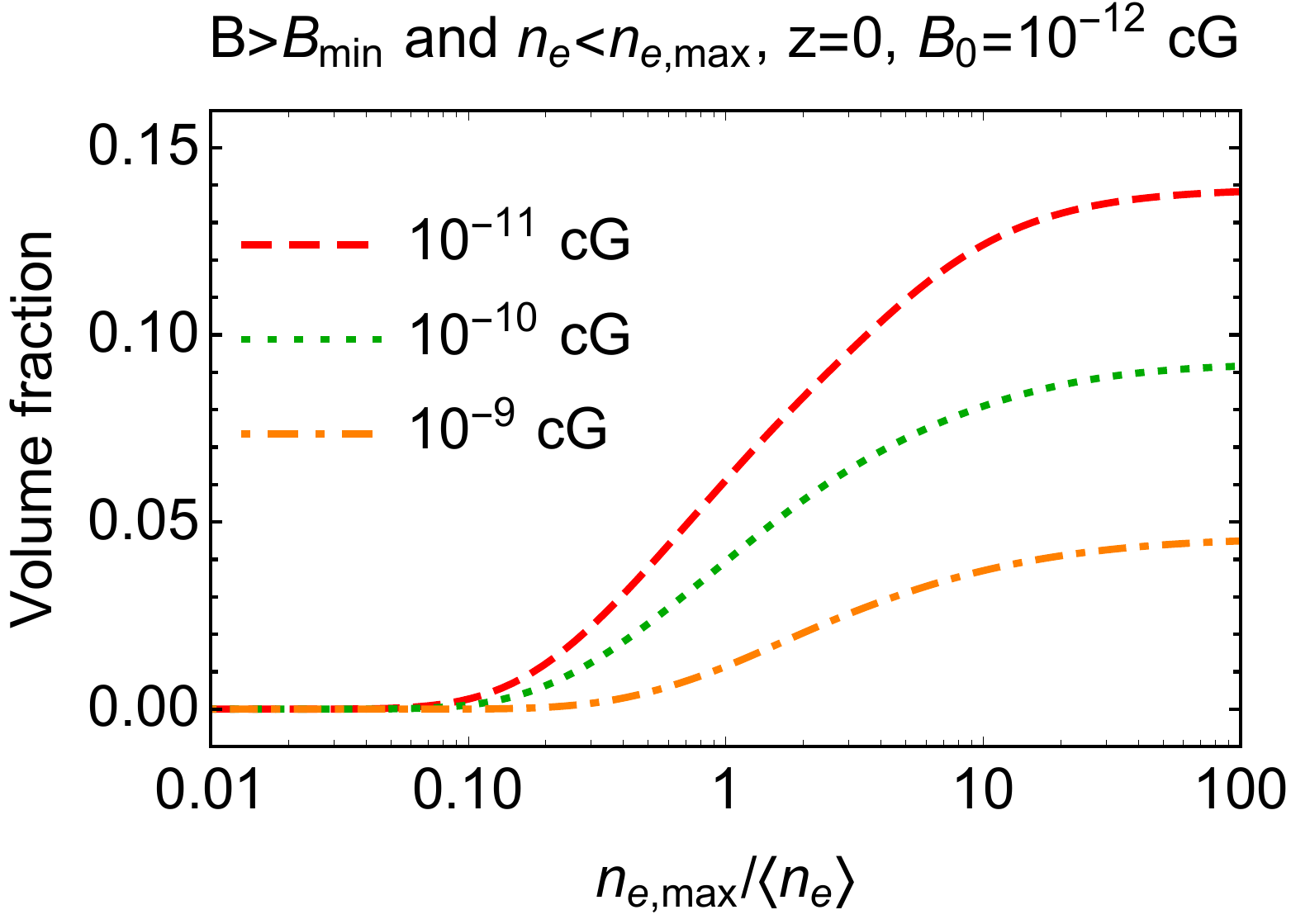}
    \includegraphics[width=0.37\textwidth]{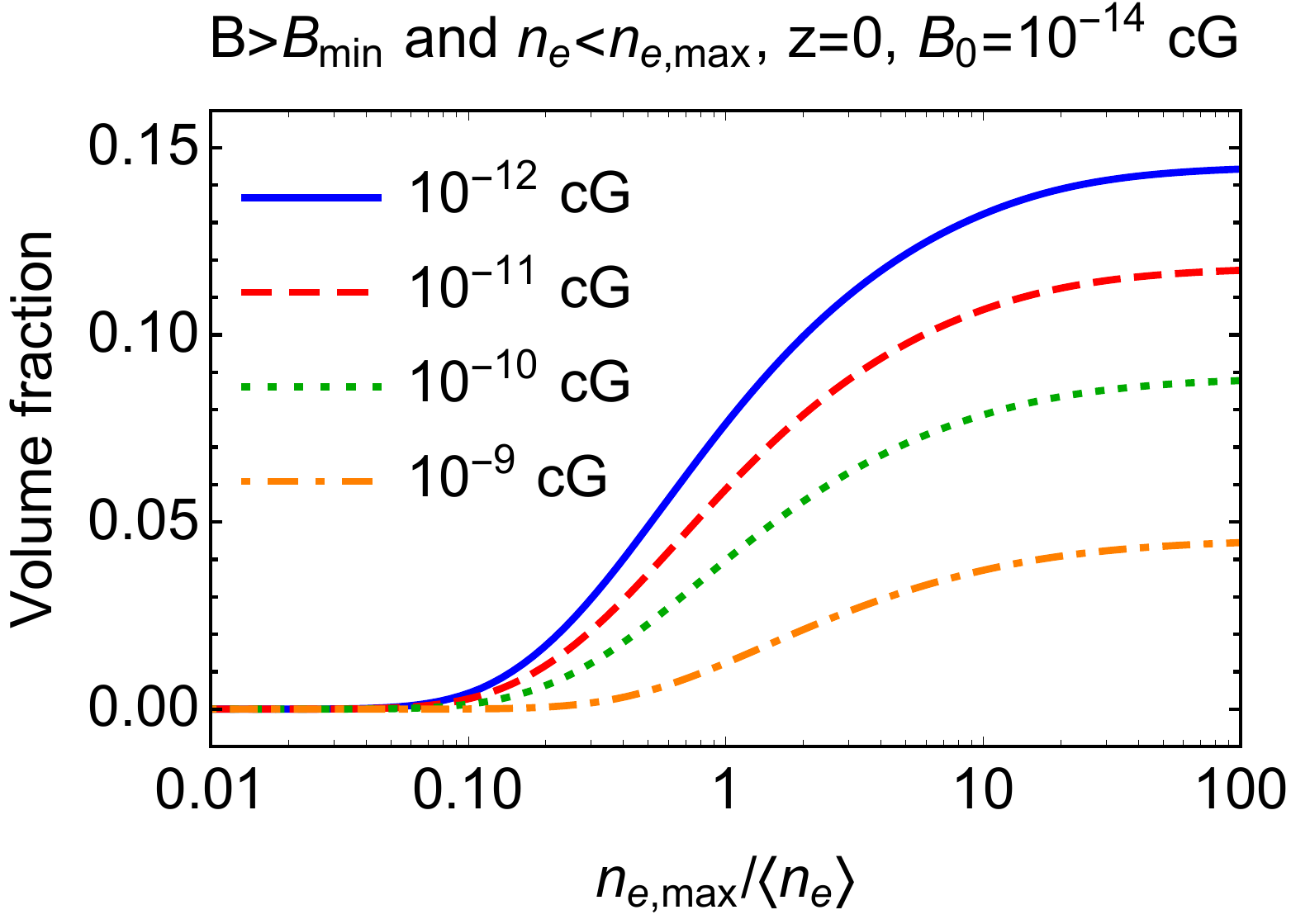} \\
    \includegraphics[width=0.37\textwidth]{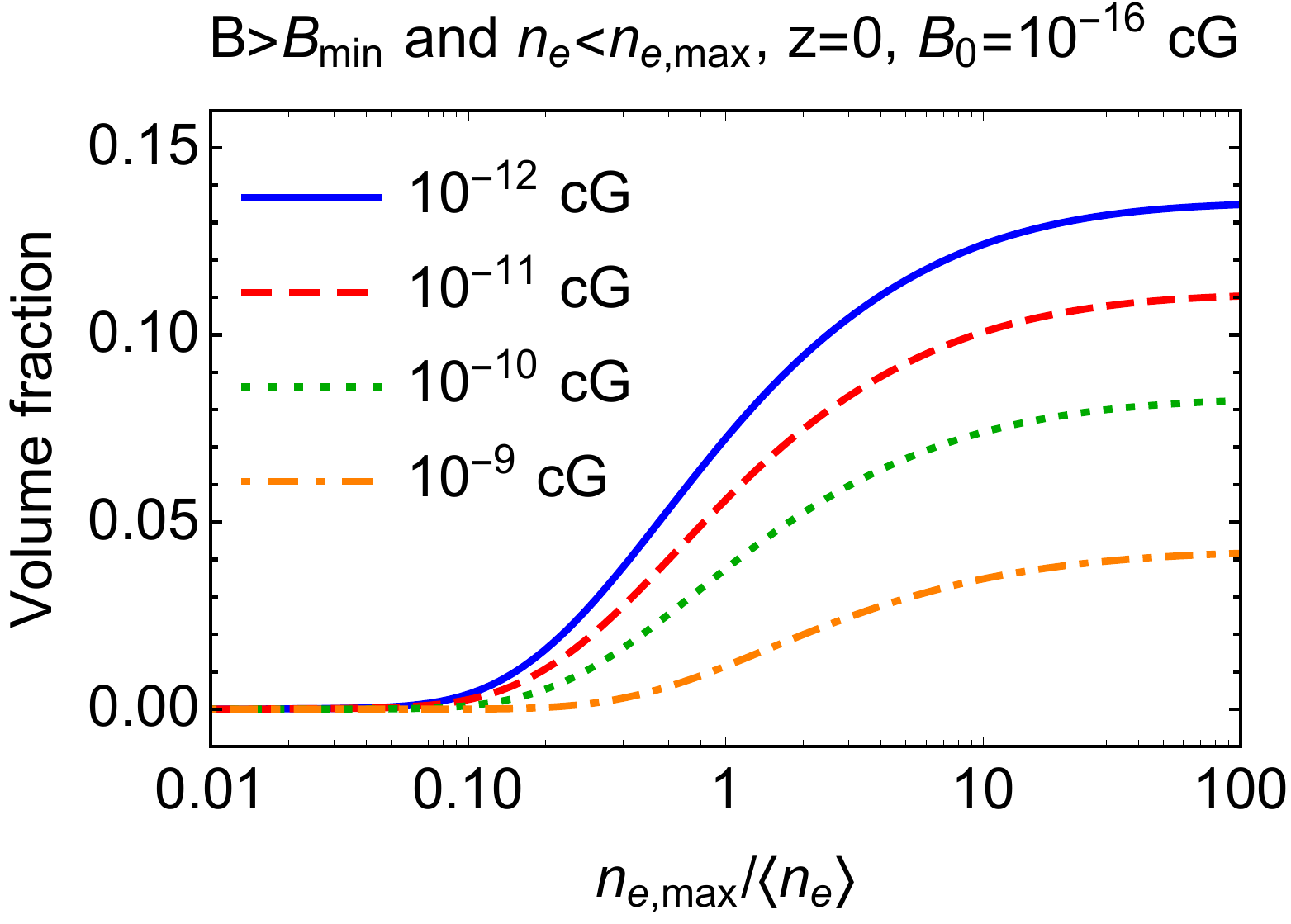}
    \includegraphics[width=0.37\textwidth]{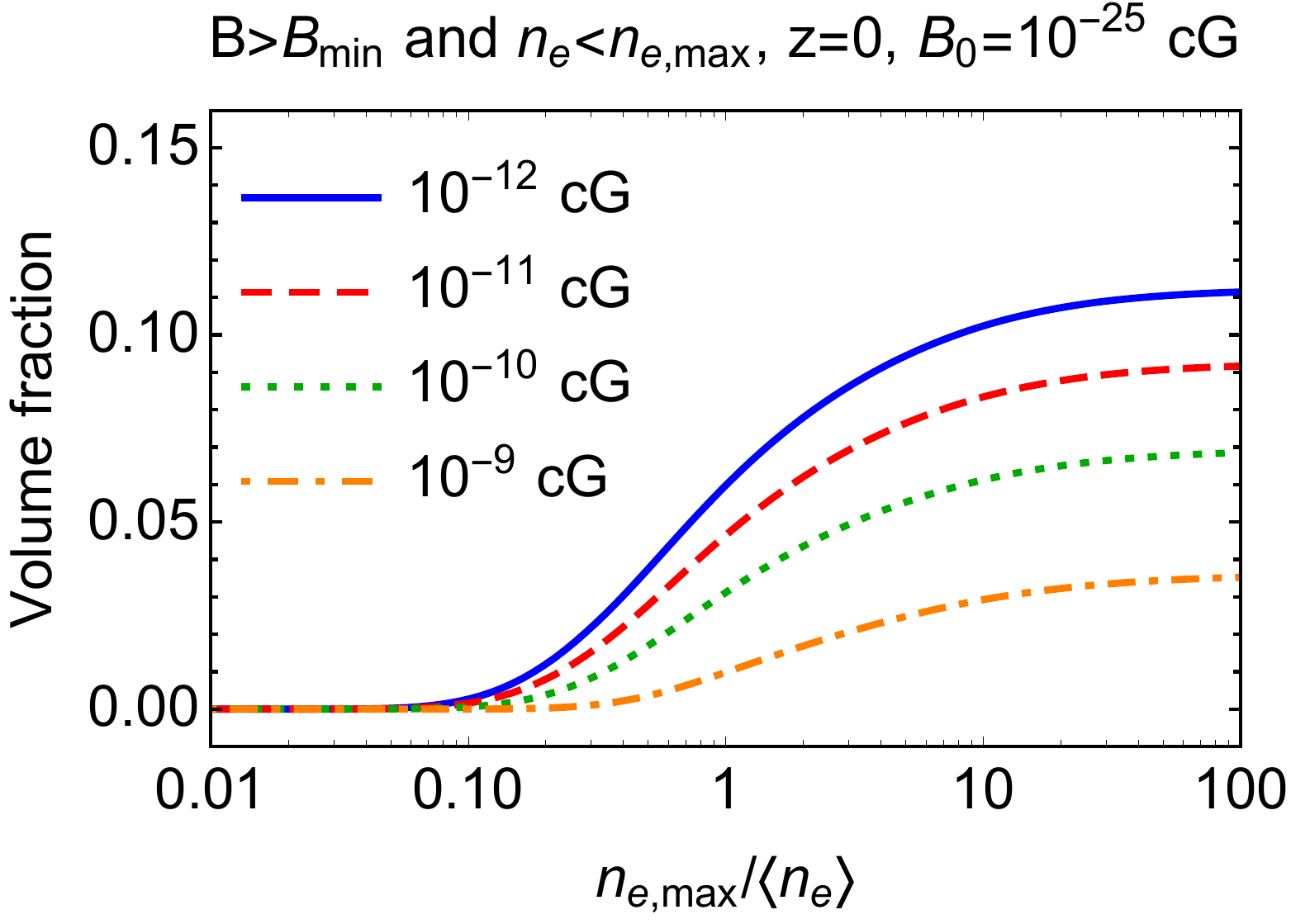}
    \caption{\textcolor{red}{Volume fractions of the regions where the magnetic field is larger than $B_{\min}$ (legend) and the electron number density is smaller than $n_{e,\max}$ (x-axis) for $z=0$ for different values of seed field: $10^{-12}$~cG, $10^{-14}$~cG, $10^{-16}$~cG, and $10^{-25}$~cG. To produce these figures we used all gas cells in each of the simulation volumes.}}
    \label{fig:25Mpc_vfrac}
\end{figure*}
  
In Fig.~\ref{fig:25Mpc_B_ne} we present for these simulations the analogous result as in Fig.~\ref{fig:ne_B} for $z=0$. The initial conditions of the magnetic field seed appear to affect mainly the lower branch while the upper branch occupies magnetic field values around $10^{-9}~\text{G}$ which is in concordance with our result presented in Fig.~ \ref{fig:B3}. In Fig.~\ref{fig:25Mpc_vfrac} we show the volume fractions for each variation as seen in Fig.~\ref{fig:volume-fraction}. \textcolor{red}{We see that the volume filling fraction weakly depends on the seed magnetic field value, and the results for the 25 cMpc/h volume are in good agreement with the fiducial TNG100 simulation adopted throughout.}

\section{Evolution of magnetic field strength with redshift}
\label{app:100Mpc}

\begin{figure*}
    \centering
    \includegraphics[width=0.48\textwidth]{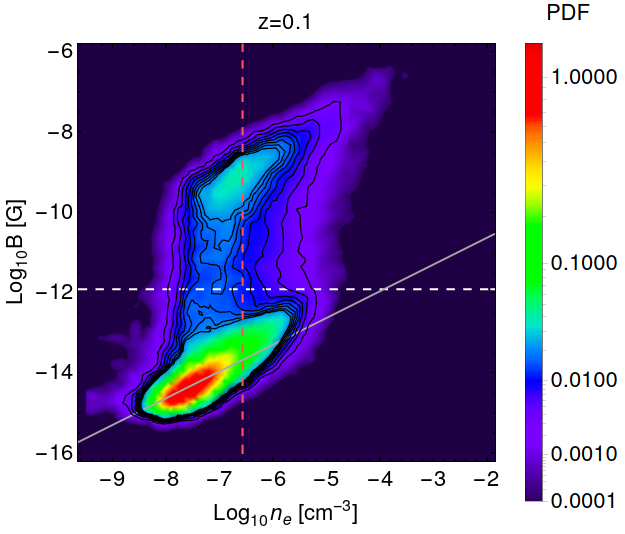}
    \includegraphics[width=0.48\textwidth]{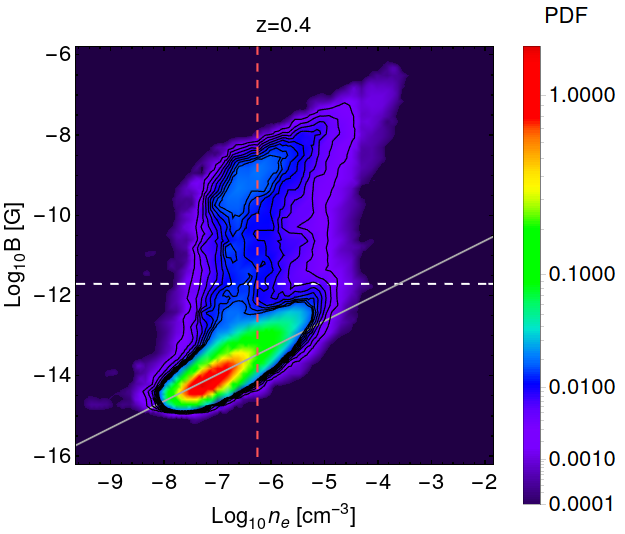}
    \includegraphics[width=0.48\textwidth]{Plots/ne_B_z1}
    \includegraphics[width=0.48\textwidth]{Plots/ne_B_z2}
    \includegraphics[width=0.48\textwidth]{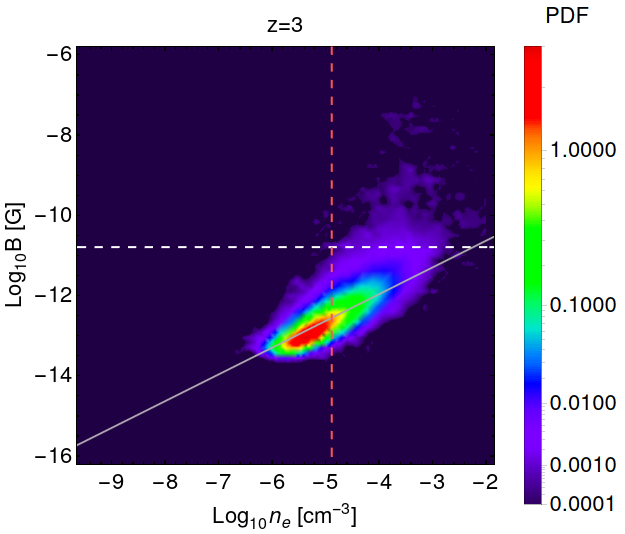}
    \includegraphics[width=0.48\textwidth]{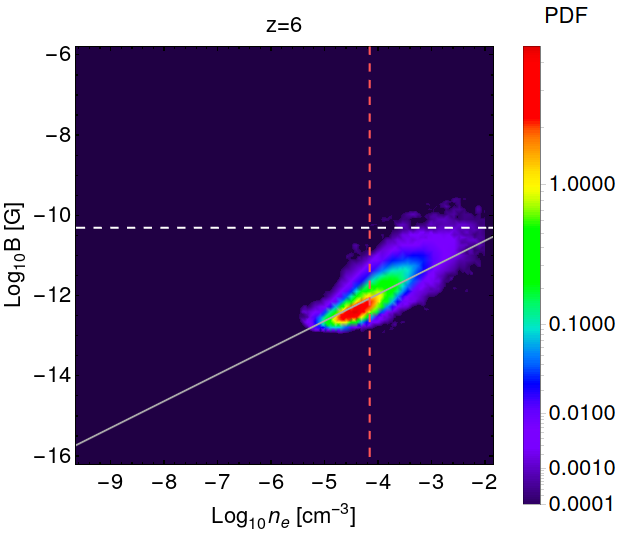}
    \caption{
    \textcolor{red}{Distribution of the magnetic field magnitude and electron number density in the TNG100 simulation using data along 200 random lines of sight in the box}
    at redshifts $z=0.1$, $0.4$, $1$, $2$, $3$, and $6$. The \textcolor{red}{white dashed} line corresponds to the comoving magnetic field value $10^{-12}$~cG that we use as smallest values of the outflow-generated magnetic field in this work. The \textcolor{red}{red dashed} line represents the average electron number density at a given redshift. \textcolor{red}{The gray dashed line show a power law $B\propto n_e^{2/3}$, that should work for the adiabatic evolution.} The seed field is $B_0 = 10^{-14}$~cG.}
    \label{fig:ne_B2}
\end{figure*}

\begin{figure*}
    \centering
    \includegraphics[width=0.48\textwidth]{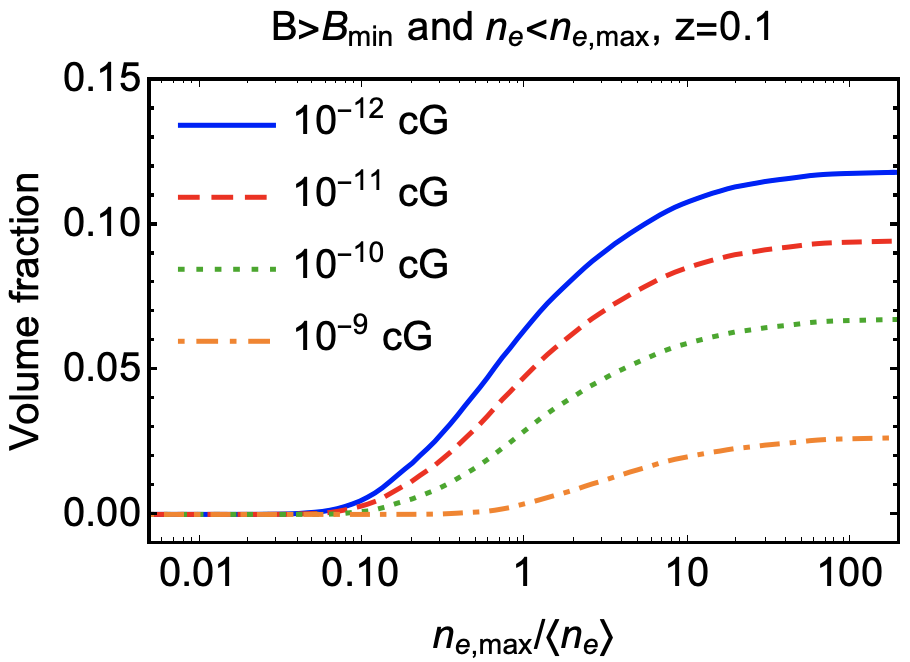}
    \includegraphics[width=0.48\textwidth]{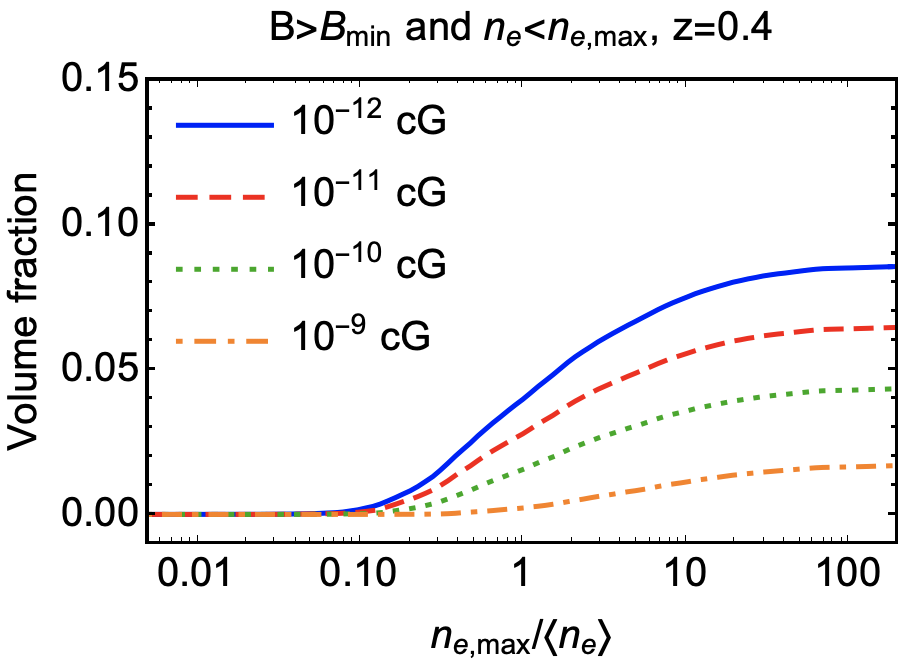}
    \includegraphics[width=0.48\textwidth]{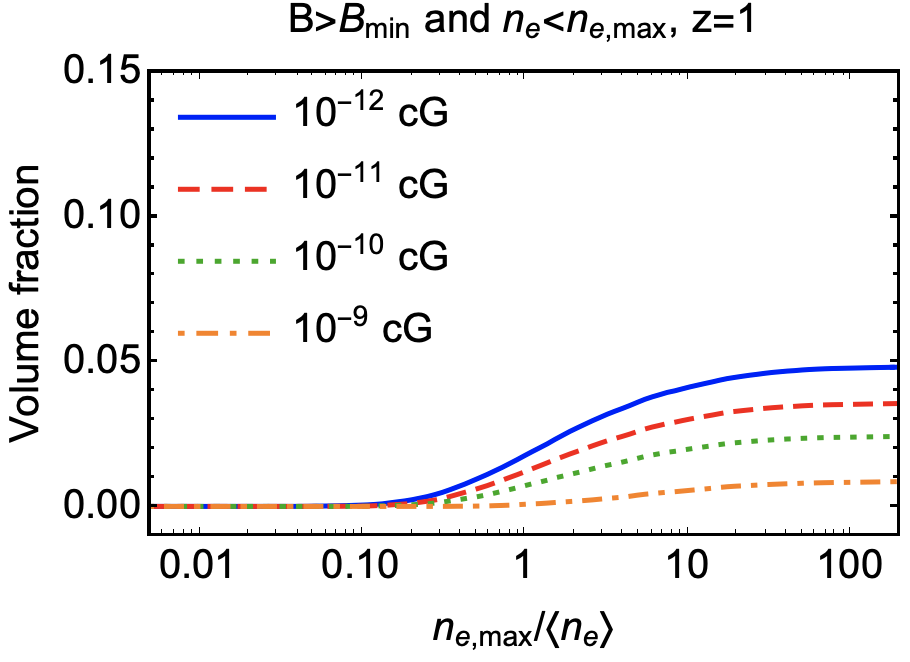}
    \includegraphics[width=0.48\textwidth]{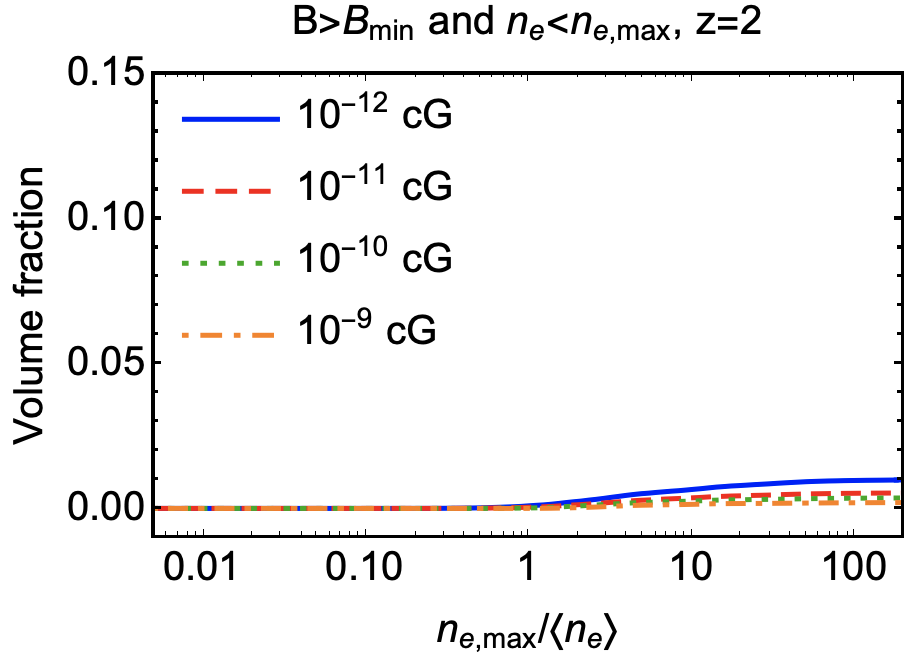}
    \includegraphics[width=0.48\textwidth]{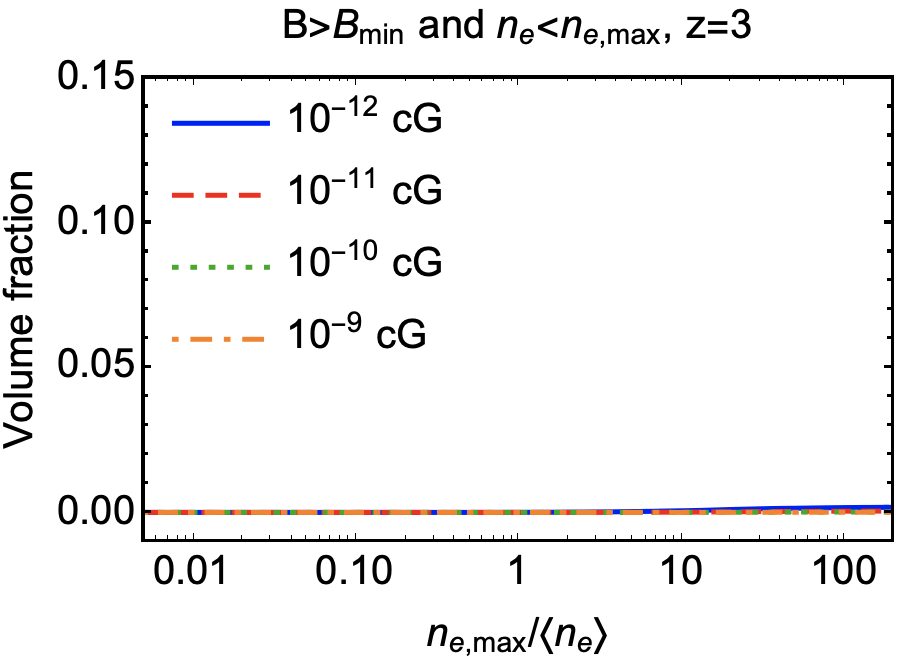}
    \includegraphics[width=0.48\textwidth]{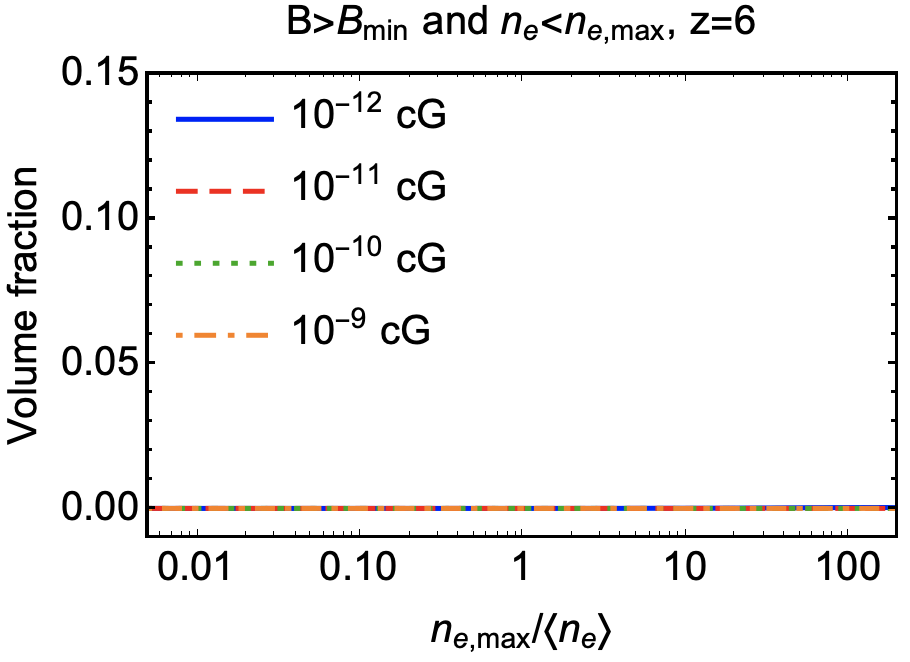}
    \caption{Volume fractions of the regions where magnetic field is larger than $B_{\min}$ and electron number density is smaller than $n_{e,\max}$ as a function of redshift (different panels), from $z=0.1$ to $z=6$ \textcolor{red}{using data along 200 random lines of sight in the TNG100 box}. The seed field is $B_0 = 10^{-14}$~cG.}
    \label{fig:volume-fraction3}
\end{figure*}

We present in Fig.~\ref{fig:ne_B2} the time progression for the correlation of electron number density and the magnitude of the magnetic field for the TNG100 simulation. We can observe how the upper branch develops with time as the universe evolves and becomes clearly distinguishable for $z\lesssim 1$. Similarly, Fig.~\ref{fig:volume-fraction3} shows the evolution of volume filling fraction for large values of magnetic fields.

\begin{figure*}
    \centering
    \includegraphics[width=0.48\textwidth]{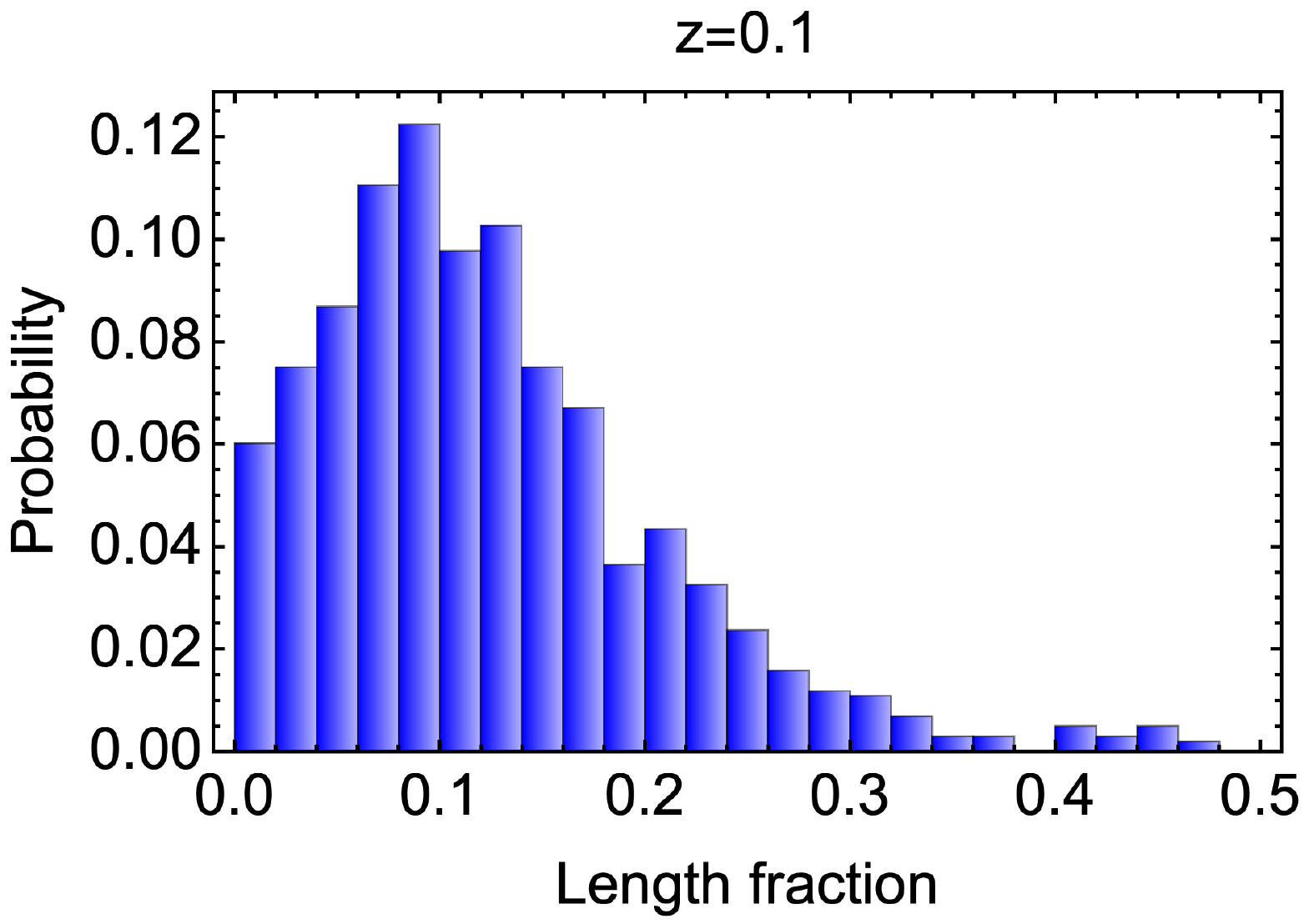}
    \includegraphics[width=0.48\textwidth]{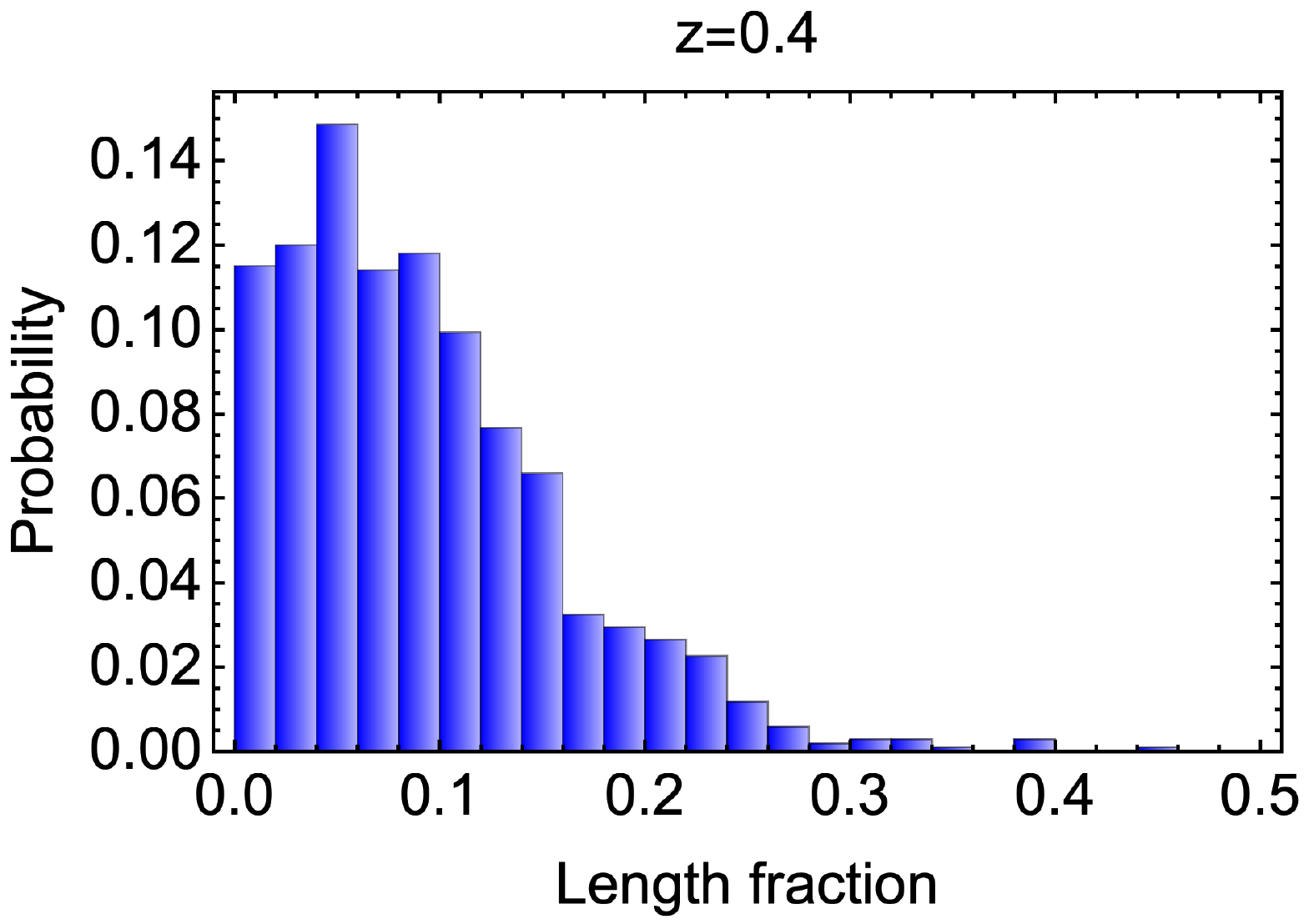}
    \includegraphics[width=0.48\textwidth]{Plots/Lfraction_z1.pdf}
    \includegraphics[width=0.48\textwidth]{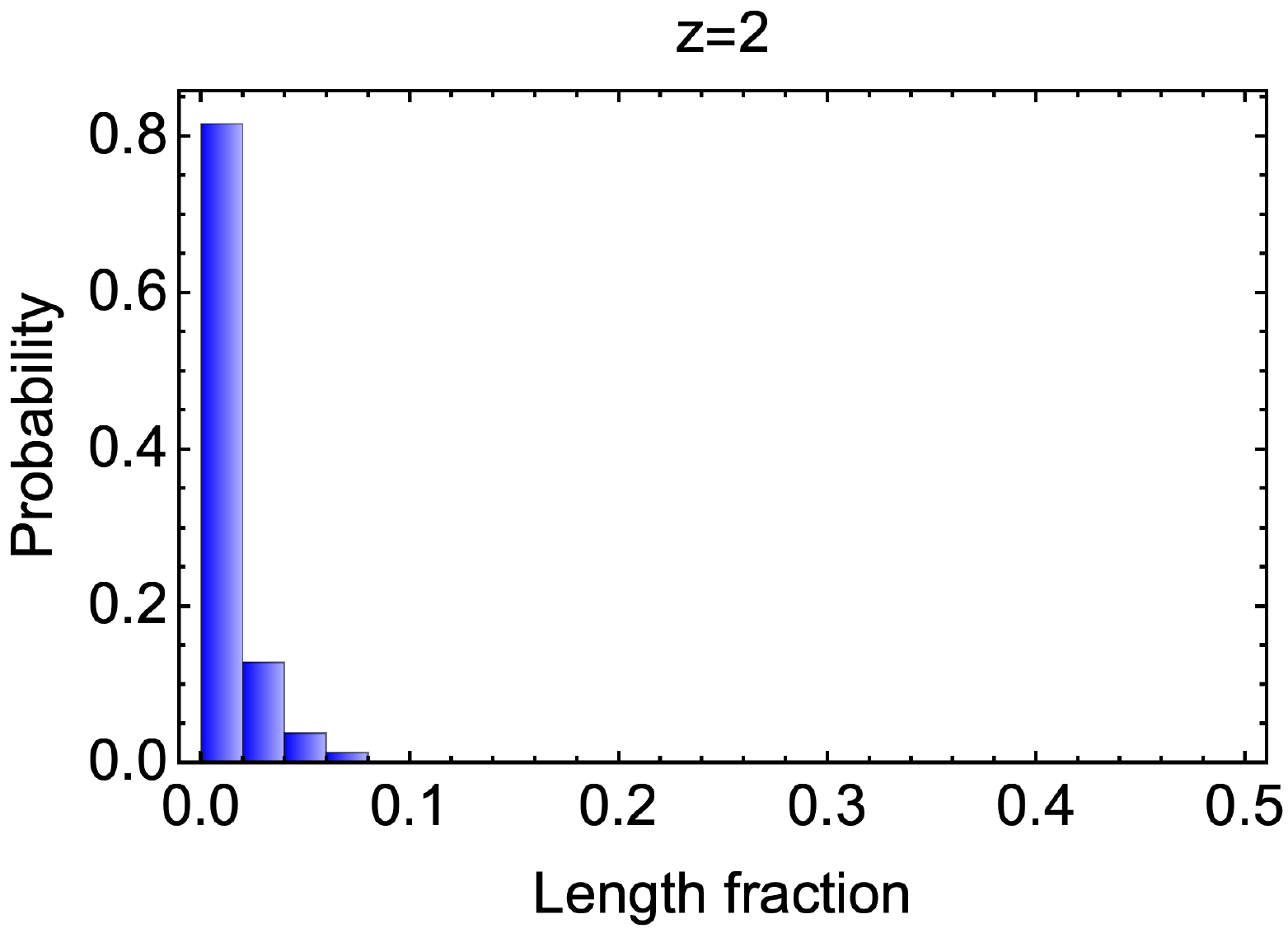}
    \includegraphics[width=0.48\textwidth]{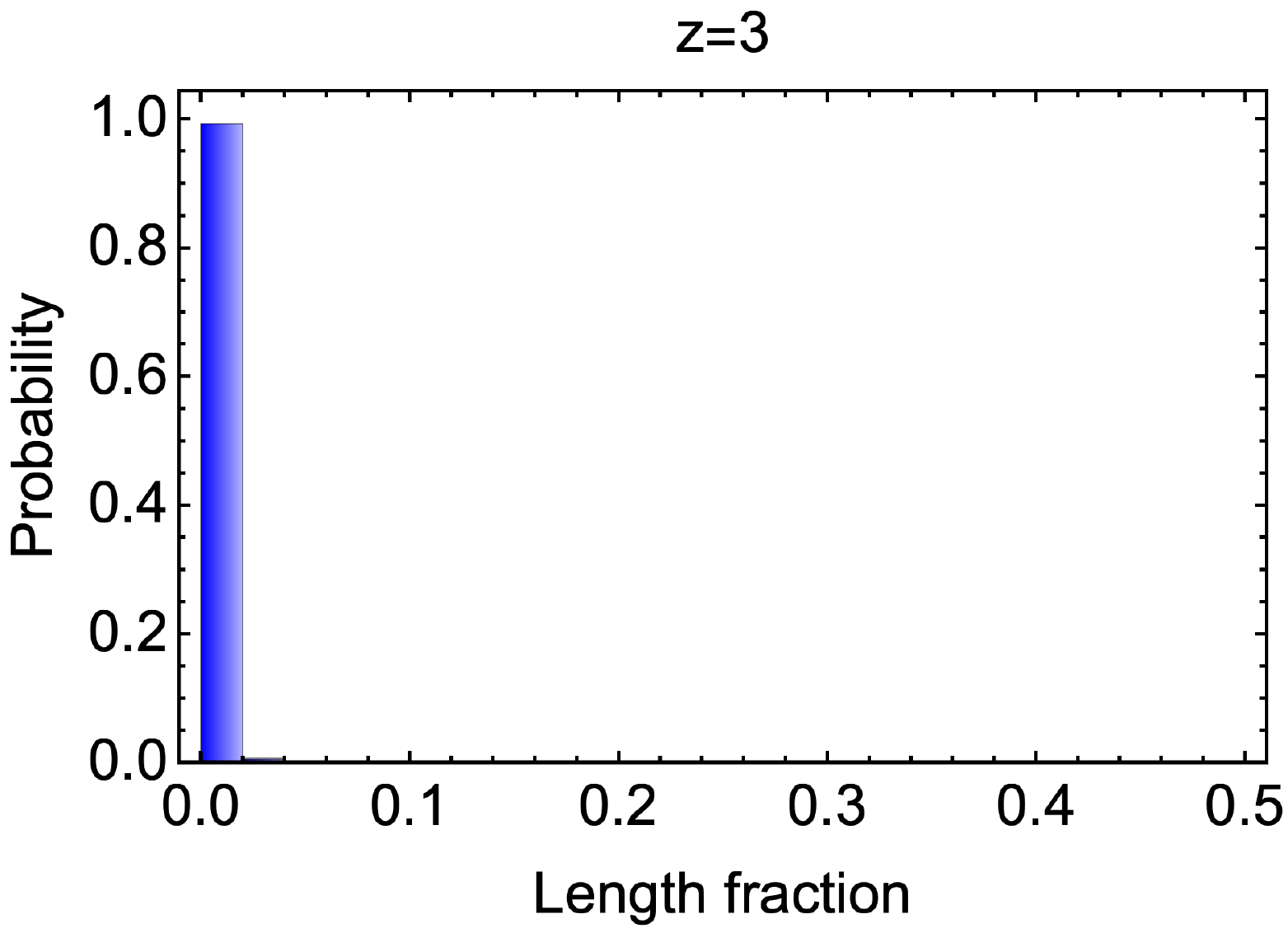}
    \includegraphics[width=0.48\textwidth]{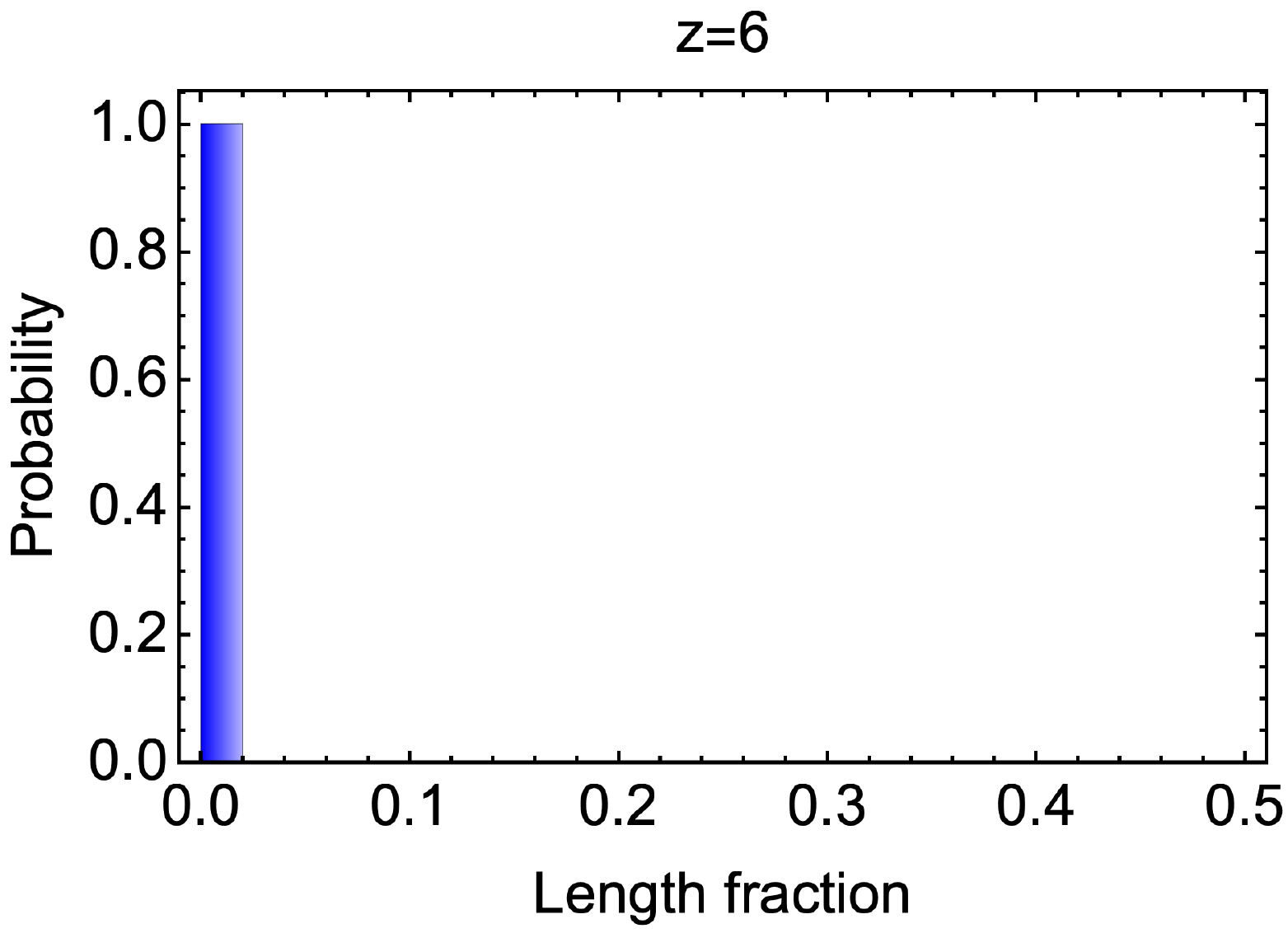}
    \caption{\textcolor{red}{Probability to find a fractional length along the line of sight with magnetic field larger than $10^{-12}$ comoving Gauss} for the 100 Mpc box at redshifts from $z=0.1$ to $z=6$. At each figure 1000 random lines of sight were taken from the simulation.}
    \label{fig:length-fraction3}
\end{figure*}

Fig.~\ref{fig:length-fraction3} displays the evolution of the length fraction of the LOS that has a magnitude of a magnetic field larger than $10^{-12}~\text{cG}$. We see that the distribution starts to broaden circa $z\approx 2$.

\section{TNG300: Dependence of the magnetic field on box size}
\label{app:300Mpc}

\begin{figure}
    \centering
    \includegraphics[width=0.48\textwidth]{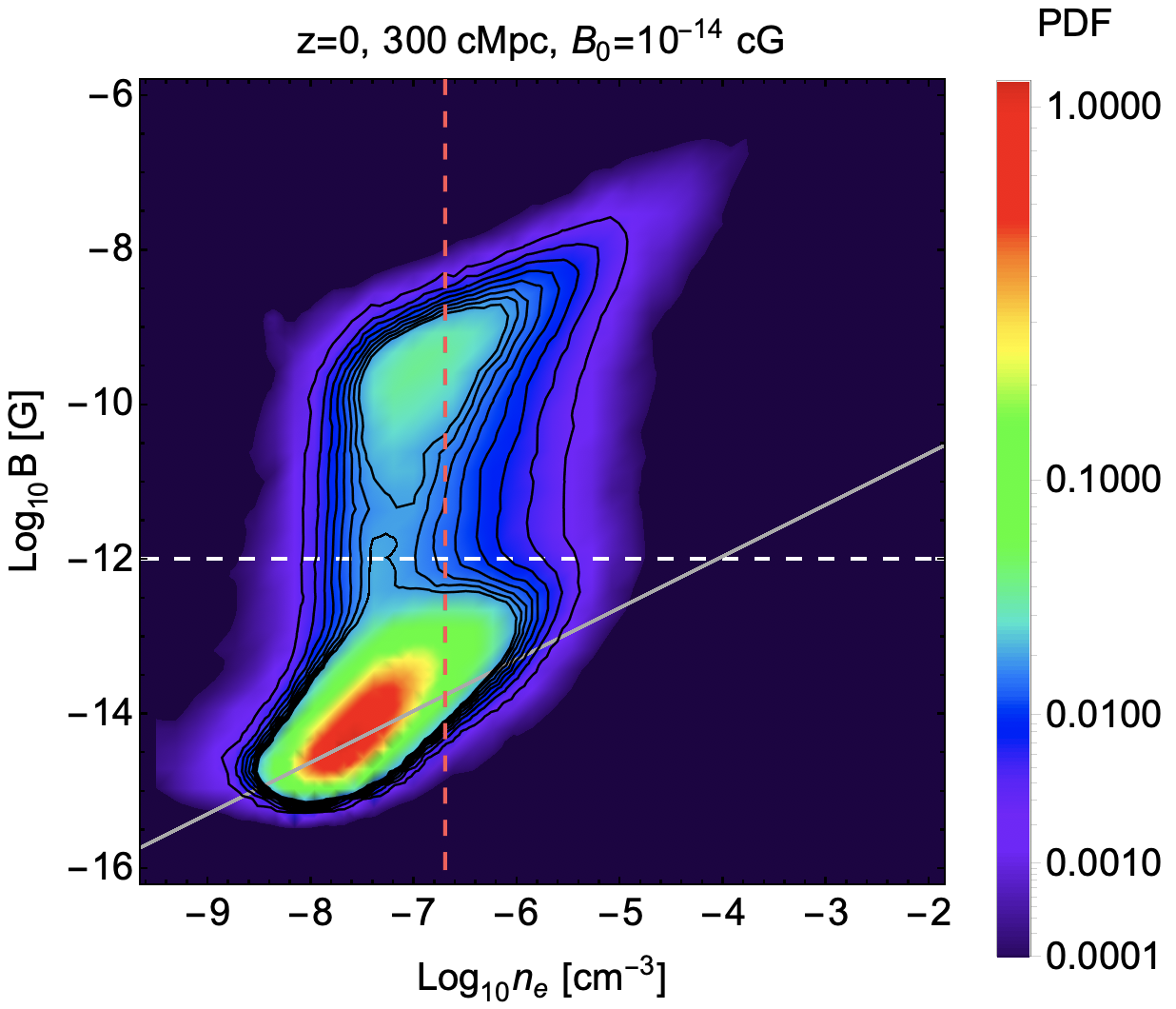}
    \caption{\textcolor{red}{Distribution of the magnetic field strength and electron number density in the TNG300 simulation using data along 420 random lines of sight in the simulation box at $z=0$, with the seed field of $B_0 = 10^{-14}$~cG. The dashed white line corresponds to the comoving magnetic field value $10^{-12}$~cG that we use as the smallest value of outflow-generated magnetic fields in this work. The red dashed line represents the average electron number density at a given redshift. The gray dashed line show a power law $B\propto n_e^{2/3}$ for the adiabatic evolution.}}
    \label{fig:ne_B3}
\end{figure}

\begin{figure*}
    \centering
    \includegraphics[width=0.48\textwidth]{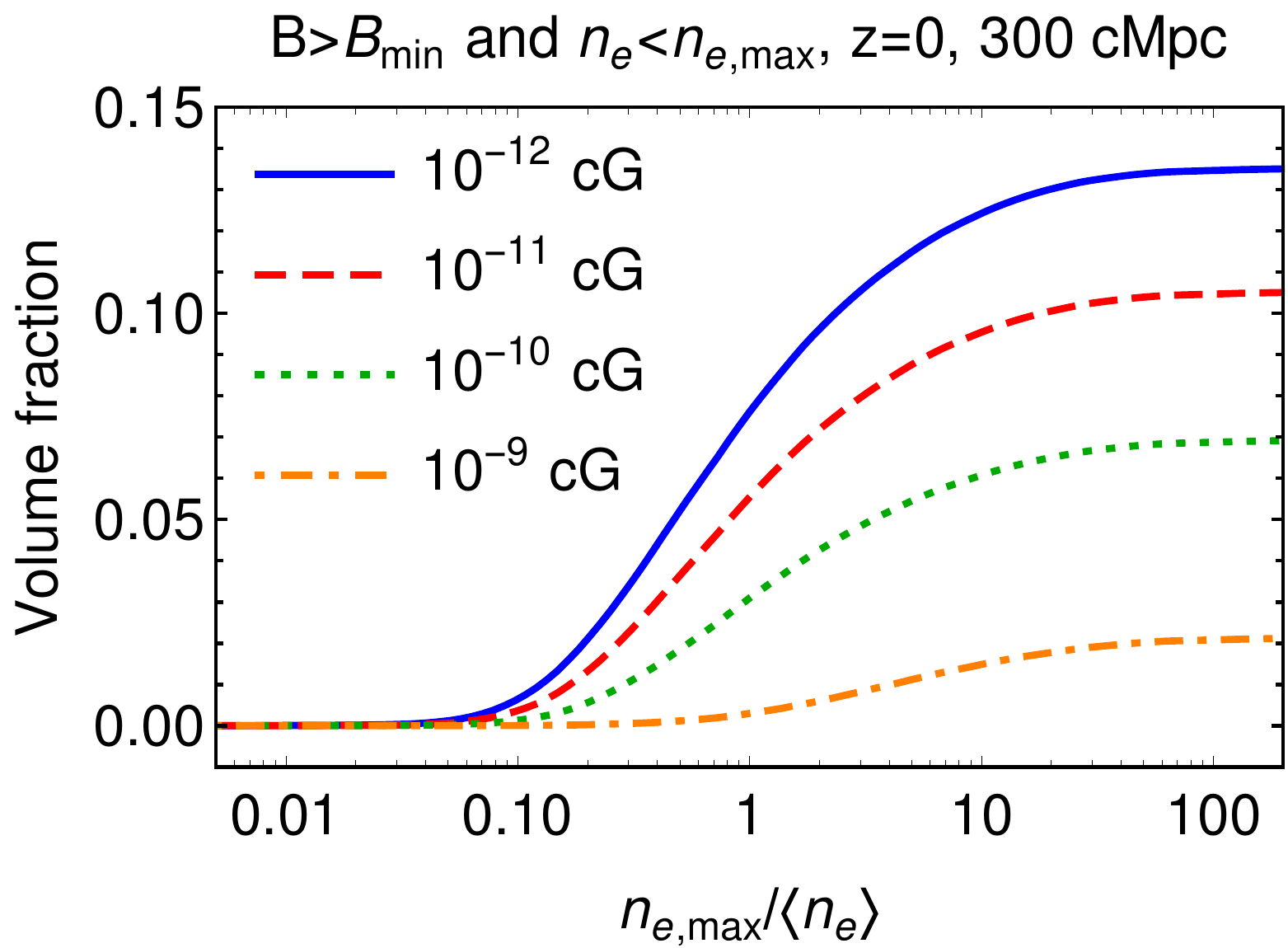}
    \includegraphics[width=0.48\textwidth]{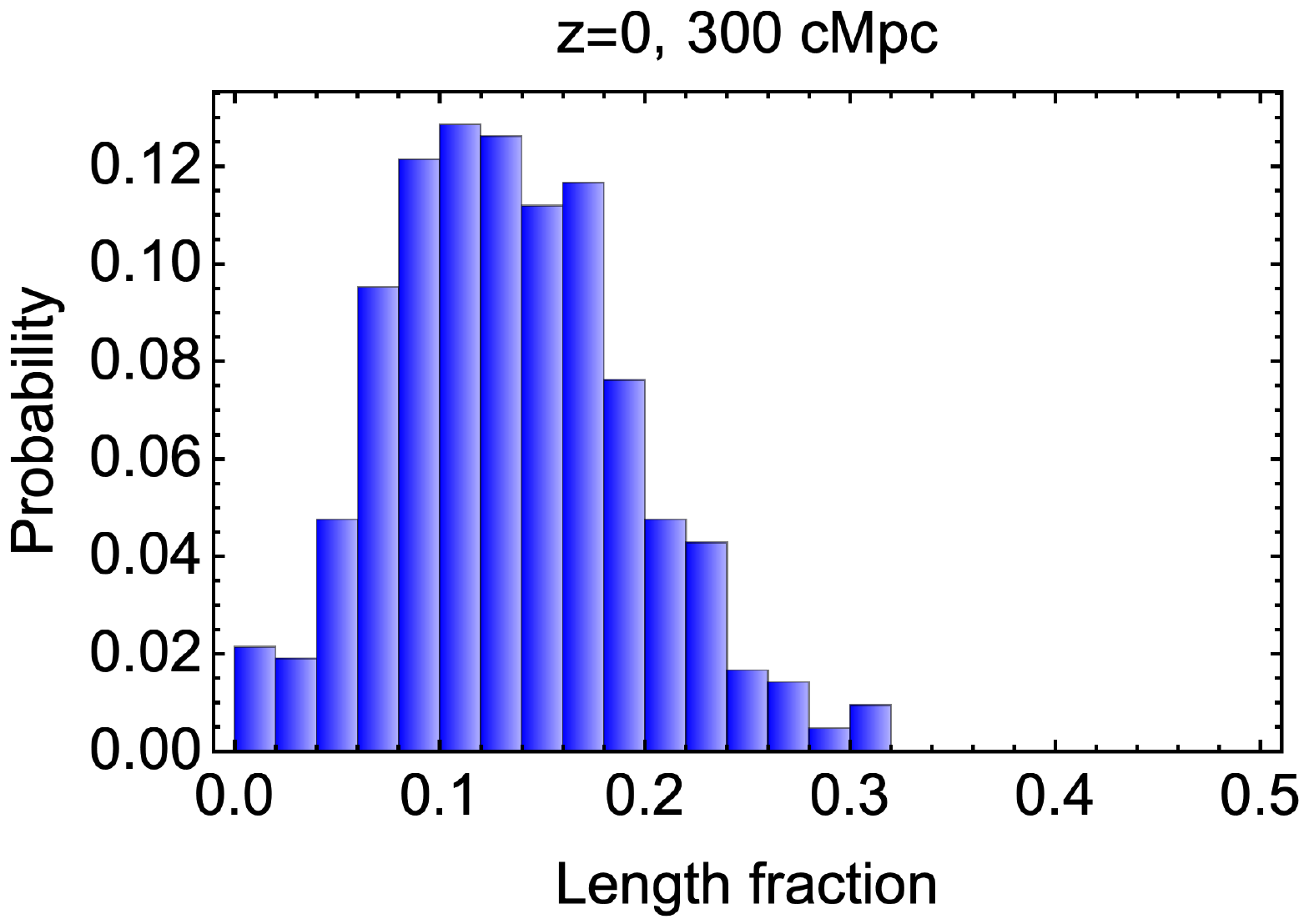}
    \caption{\textit{Left panel}: volume fractions of the regions where magnetic field is larger than $B_{\min}$ and electron number density is smaller than $n_{e,\max}$ for $z=0$ in the 300~cMpc box. The seed field is $B_0 = 10^{-14}$~cG, \textcolor{red}{as in TNG100}. \textit{Right panel}: \textcolor{red}{Probability to find a given fractional length along the line of sight with magnetic field larger than $10^{-12}$ comoving Gauss} for the 300 Mpc box at redshift $z=0$, based on 420 random lines of sight.}
    \label{fig:volume-fraction2}
\end{figure*}

To explore the impact of the finite simulation size, we repeat our analysis on the larger TNG300 simulation, \textcolor{red}{which however is performed at} somewhat lower mass resolution \textcolor{red}{than the TNG100 box: namely, dark matter and baryonic mass resolution of} $5.9\times 10^7~\mathrm{M_\odot}$ and $1.1\times 10^7~\mathrm{M_\odot}$ respectively. In Fig.~\ref{fig:ne_B3} we present the magnitude of the magnetic field versus the electron number density. Comparing with the results for the same redshift shown in Fig.$\ref{fig:ne_B}$, we observe the same bimodal distribution within the same value ranges for both the TNG100 and TNG300-1 simulations. 
\textcolor{red}{Comparing Fig.~\ref{fig:volume-fraction2} (left) and Fig.~\ref{fig:volume-fraction} we see that TNG300 has smaller volume filling fraction (for example, for $B>10^{-12}$ the volume filling fraction in TNG100 is $0.14$, while in TNG300 is $0.135$). We believe that this is an effect of poorer resolution of the TNG300 simulation in comparison to the fiducial TNG100.}

\end{document}